\documentclass[aps,showpacs]{revtex4}
\usepackage{graphicx}

\def\ket{\rangle}
\def\<{\langle}
\def\>{\rangle}

\begin{document}

\title{Construction of Elementary Gates in Quantum Computation by Joint Measurement}

\author{Jia-Qi Jin$^{1}$ and Gui-Lu Long$^{1,2}$}

\affiliation{$^1$ Key Laboratory For Quantum Information and
Measurements and Department of Physics, Tsinghua University,
Beijing 100084, P.R. China\\
$^2$ Center for Atomic and Molecular NanoSciences, Tsinghua
University, Beijing 100084, P.R. China}

\date{\today}

\begin{abstract}
In this paper, elementary quantum gate operations, such as the
phase gate, the controlled-NOT gate, the swap and the Fredkin gate
are constructed using joint measurement and pairs of entangled
qubit pairs. The relation between the state of the entangled pair
and the joint measurement basis is discussed. Some other
generalization is also discussed.
\end{abstract}

 \maketitle

\section{Introduction}
\label{s1}

 Entanglement is one important concept in quantum mechanics.
In quantum teleportation, maximally entangled states are the
pivotal resource. It is also the source of power in quantum
computation. Usually, quantum computation starts from an initial
quantum state, say, $|0...0\ket$. Quantum gate operations are
unitary operations that are constructed using a finite set of
basic universal quantum computing gates\cite{barenco}. At the end
of a computation, a measurement is performed on the register so
that the outcome is read out. Many schemes have been proposed, and
considerable progress has been made over the last decade. However,
it is still a daunting task to build a practical quantum computer.
Though it is difficult to build complicated quantum system and to
perform complex computation operations, small quantum systems are
already easily available and some basic quantum operations can be
performed. Recently there have been proposals to perform
complicated quantum computation with small quantum systems such as
entangled pairs of qubits and simple measurement. It has been
shown that quantum computation can be performed using single qubit
operations, joint two-qubit measurement and together with some
simple entangled quantum states such as
GHZ-states\cite{gottesman,knill,nielsen1,nielsen2,leung}. In these
schemes, the initial state of the quantum computer may be a
separable state and this spares the need to perform complicated
gate operations, however at the expense of using some number of
entangled states such as Einstein-Podolsky-Rosen (EPR) pairs.
Alternatively, quantum computation can be performed by starting
from a very complicated entangled states, and then proceeding to
the end by merely measurements. The cluster state quantum
computation scheme\cite{briegel,briegel2} is typical example of
these schemes. Valence bond state,  proposed by Ian Affleck $et\;
al.$\cite{vbs0} and used in condensed matter
physics\cite{Affleck,Fannes,Zhang,Nakamura,Koretsune,Zhang2}, has
also been found to play an important role in quantum computation
recently\cite{cirac1,cirac2,vbs}. Verstraete and Cirac\cite{vbs}
proved that the cluster state quantum computation and
valance-bond-state computation are equivalent.

In Ref.\cite{vbs}, it was shown that universal quantum computation
can be carried out by using only pairs of qubits in singlet state
and by performing joint two-qubit and three-qubit measurement. It
has the advantage of being deterministic and using a simple
initial product state. Single-qubit operation can be implemented
by performing a Bell-basis-like measurement on the target qubit
and another qubit from a singlet. Controlled-Z gate is implemented
by performing two GHZ-like state measurements on the two qubits
and three singlets system. These two unitary gate operations are
the basic building blocks for more complicated gate operations. As
this scheme is practically appealing, it is interesting to
construct the elementary gate operations in this scheme. In this
paper, we will provide the constructions of the elementary gate
operations for this scheme. The paper is organized as follows. In
section \ref{s2}, we briefly summarize the construction of basic
gates of Ref.\cite{vbs}. In section \ref{s3}, we give the results
for the well-known elementary gates mentioned in\cite{nielsen3},
which are the basic ingredients for quantum algorithms in quantum
computation. In section \ref{s4}, we give a brief summary.

\section{The basic gates}
\label{s2}

Here we briefly review the basic gate operation in the scheme in
Ref.\cite{vbs}. In Ref.\cite{vbs}, basic single-qubit  local
unitary operation on a particle can be implemented in a
teleportation fashion. Suppose the state of particle $A$ is
$|\psi\ket$, and particles $B$ and $C$ are in an entangled state
$|H\ket$. By performing a joint measurement on qubits A and B in
the following-basis
\begin{eqnarray}
|\alpha\rangle=(U^+\sigma_\alpha\otimes I)|H\rangle, \;\;(\alpha
=1,2,3,4), \label{e1}
\end{eqnarray}
where $\sigma_\alpha$ denote the Pauli matrices $(\sigma_4=I)$ and
\begin{eqnarray}
|H\rangle&=&(|00\rangle+|01\rangle+|10\rangle-|11\rangle)/2\nonumber\\
         &=&(|0+\ket+|1-\ket)/\sqrt{2},
\label{eh}
\end{eqnarray}
is a maximally entangled state, and
\begin{eqnarray}
|\pm\ket=(|0\ket\pm|1\ket)/\sqrt{2}\end{eqnarray} is the
eigenstates of $\sigma_x$. The state of particle $C$ becomes
$\sigma_\alpha U|\psi\ket$ where $\alpha$ is the outcome of the
measurement in basis $\{|\alpha \ket, \alpha=1, 2, 3, 4.\}$.
Particle $C$ then works as the role of particle in the following
process of the quantum computation.

It is worth mentioning here that the entangled state $|H\ket$ can
be replaced by arbitrary maximally entangled states such as the
Bell-basis states,
$|\phi^{\pm}\ket=(|00\rangle\pm|11\rangle)/\sqrt{2} $ and
$|\psi^{\pm}\ket=(|01\rangle\pm|10\rangle)/\sqrt{2} $. However
this substitution can not be used for the controlled-Z gate.

\begin{figure}
\begin{center}
\includegraphics[width=7cm,angle=0]{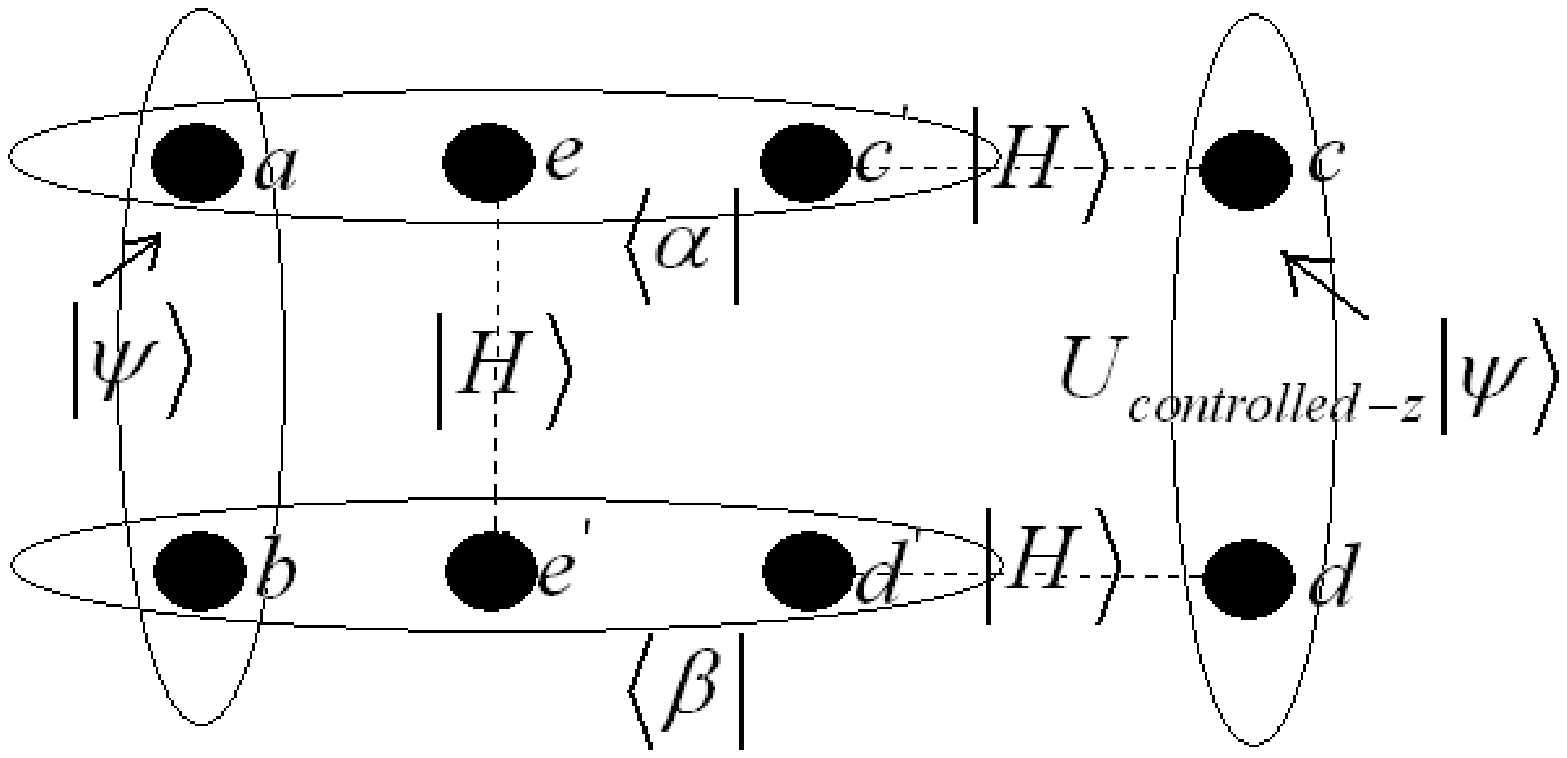}
\end{center}
\caption{ Implementation of the controlled-phase gate in
Ref.\cite{vbs}.} \label{f2a}
\end{figure}

 The controlled-Z gate,
\begin{eqnarray}
U_{ctr-z}=|00\rangle\langle00|+|01\rangle\langle01|+|10\rangle\langle10|-|11\rangle\langle11|
\label{e2}
\end{eqnarray}
proposed in Ref.\cite{vbs} involves two GHZ-like state joint
measurements as redrawn in Fig.\ref{f2a} where we have given
explicit labelling to the eight qubits involved.

By doing three-qubit joint measurement of particles on $a$, $e$,
$c^{\prime}$ and $b$, $e^{\prime}$, $d^{\prime}$ in the following
basis,
\begin{eqnarray}
\{|\alpha\rangle\}=\{|\beta\rangle\}=\{(\sigma_x)^i\otimes(\sigma_x)^j\otimes
I\;(|000\rangle\pm|111\rangle) \},\label{e3}
\end{eqnarray}
where $i,j\in {\{0,1\}}$, the controlled-Z gate is implemented
between qubits $c$ and $d$. The singlet states of the entangled
pairs between $e$ and $e^{\prime}$ , $c$ and $c^{\prime}$, $d$ and
$d^{\prime}$ are all $|H\rangle$.

\begin{figure}
\begin{center}
\includegraphics[width=7cm,angle=0]{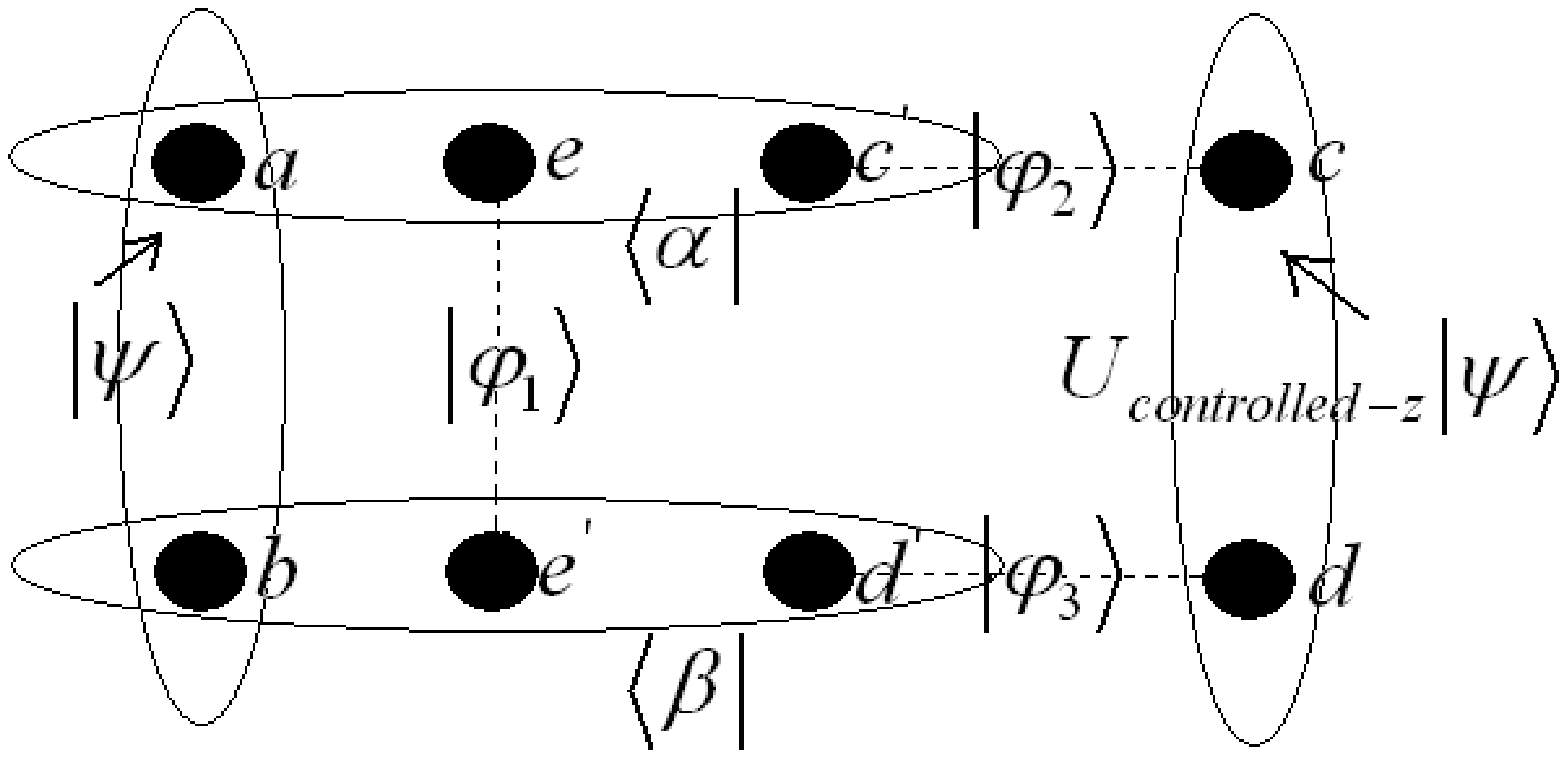}
\end{center}
\caption{ The implementation of the controlled-phase gate using
different pairs of entangled states.} \label{f2b}
\end{figure}

By replacing the singlet state $|H\ket$ by other entangled states,
it is also possible to construct the gate. The detailed
combination of the states used for each pair and the appropriate
basis for the joint measurement are given in Table \ref{t1}. The
main conclusion of this table is that the measuring-basis is
determined by the state of the $e\; e'$ pair, the states of
$c\;c'$ and $d\;d'$ pairs are irrelevant. For example, as in
Fig.\ref{f2b}, assuming the state of the two original qubits is
\begin{eqnarray}
|\psi\rangle_{ab}=c_0|00\rangle+c_1|01\rangle+c_2|10\rangle+c_3|11\rangle),
\label{e4}
\end{eqnarray}
and the states of the $(e\;e')$, $(c\;c')$ and $(d\;d')$ pairs
denoted by $|\varphi_1\ket$, $|\varphi_2\ket$ and $|\varphi_3\ket$
respectively, are in the Bell-basis states,
\begin{eqnarray}
|\varphi_1\rangle=|\varphi_2\rangle=|\varphi_3\rangle=\frac{1}{\sqrt{2}}(|00\rangle+|11\rangle).\label{e5}
\end{eqnarray}
If the measuring bases of Eq.(\ref{e3}) were used,  half of the
information in $(c_0,c_1,c_2,c_3)$ would be lost. For instance, if
$|\alpha\rangle=|\beta\rangle=|000\rangle+|111\rangle$, then the
wave function of $c$ and $d$ would become
$c_0|00\rangle+c_3|11\rangle$, which have lost state information
in this process due to the form of entangled state of $e$ and
$e^{\prime}$. To see this more clearly, we write the state of the
eight-qubit system of Fig.\ref{f2b} as $|\Phi\ket$,
\begin{eqnarray}
|\Phi\rangle=|\Phi_1\rangle+|\Phi_2\rangle=|\psi\rangle\otimes|H
\rangle_{ee^{\prime}}\otimes|\varphi_2\rangle\otimes|\varphi_3\rangle,\label{e6}
\end{eqnarray}
where
\begin{eqnarray}
|\Phi_1\rangle&=&|\psi\rangle\otimes\frac{1}{\sqrt{2}}|0_e+_{e^{\prime}}\rangle\otimes
|\varphi_2\rangle\otimes|\varphi_3\rangle,\label{e7}\\
|\Phi_2\rangle&=&|\psi\rangle\otimes\frac{1}{\sqrt{2}}|1_e-_{e^{\prime}}\rangle\otimes
|\varphi_2\rangle\otimes|\varphi_3\rangle. \label{e8}
\end{eqnarray}
If the entangled state between $e$ and $e^{\prime}$ is the
$|H\ket$ state, then the three-qubit measurement projections
$\langle0_b0_{e^{\prime}}0_{d^{\prime}}|\Phi_1\rangle$,
$\langle0_b0_{e^{\prime}}0_{d^{\prime}}|\Phi_2\rangle$,
$\langle1_b1_{e^{\prime}}1_{d^{\prime}}|\Phi_1\rangle$,
$\langle1_b1_{e^{\prime}}1_{d^{\prime}}|\Phi_2\rangle$ are all
none-zero. However if we change the entangled state of $e$ and
$e^{\prime}$ to state
$\frac{1}{\sqrt{2}}(|0_e0_{e^{\prime}}\rangle+|1_e1_{e^{\prime}}\rangle)$,
and turn $|\Phi_1\rangle$,$|\Phi_2\rangle$ into
\begin{eqnarray}
|\Phi_1\rangle&=&|\psi\rangle\otimes\frac{1}{\sqrt{2}}|0_e0_{e^{\prime}}\rangle\otimes
|\varphi_2\rangle\otimes|\varphi_3\rangle,\label{e51}\\
|\Phi_2\rangle&=&|\psi\rangle\otimes\frac{1}{\sqrt{2}}|1_e1_{e^{\prime}}\rangle\otimes
|\varphi_2\rangle\otimes|\varphi_3\rangle, \label{e52}
\end{eqnarray} then  two out of four projections become zero, and this leads to
loss of components in the operation.
\begin{widetext}
\begin{center}
\begin{table}[tph]
\begin{center}
\caption{Measuring-basis for the controlled-Z gate}\label{t1}
\begin{tabular}{|l|c|c|c|r|}\hline
$|\varphi\ket_{ee'}$ & $|\varphi\ket_{cc'}$ & $|\varphi\ket_{dd'}$
& $\{|\alpha\ket_{aec'}\}$ & $\{\beta\ket_{be'd'}\}$\\ \hline
$|H\ket$ & $|\phi^{\pm}\ket$, $|\psi^{\pm}\ket$, $|H\ket$ &
$|\phi^{\pm}\ket$, $|\psi^{\pm}\ket$, $|H\ket$ &
$(\sigma_x)^i\otimes(\sigma_x)^j\otimes I(|000\ket\pm|111\ket)$ &
$(\sigma_x)^i\otimes(\sigma_x)^j\otimes I(|000\ket\pm|111\ket)$\\
\hline $|\psi^{\pm}\ket$, $|\phi^{\pm}\ket$ & $|\phi^{\pm}\ket$,
$|\psi^{\pm}\ket$, $|H\ket$ & $|\phi^{\pm}\ket$,
$|\psi^{\pm}\ket$, $|H\ket$ &
$(\sigma_x)^i\otimes(\sigma_z)^j\otimes I(|0+0\ket\pm|1-1\ket)$ &
$(\sigma_x)^i\otimes(\sigma_z)^j\otimes I(|000\ket\pm|111\ket)$\\
\hline
\end{tabular}
\end{center}
\end{table}
\end{center}
\end{widetext}

In this case,  if we change the measuring-basis turn
$\{|\alpha\rangle\}$ into
\begin{eqnarray}
\{|\alpha\rangle\}=\{(\sigma_x)^i\otimes(\sigma_z)^j
\otimes\;I(|0_a+_e0_{c^{\prime}}\rangle\pm|1_a-_e1_{c^{\prime}}\rangle)\}
\label{e9}
\end{eqnarray}
where $(i,j\in {\{0,1\}})$ and keep the state of $e$ and
$e^{\prime}$ as
$\frac{1}{\sqrt{2}}(|0_e0_{e^{\prime}}\rangle\pm|1_e1_{e^{\prime}}\rangle)$
or
$\frac{1}{\sqrt{2}}(|0_e1_{e^{\prime}}\rangle\pm|1_e0_{e^{\prime}}\rangle)$,
and the measuring-basis  $\{|\beta\rangle\}$ as before, the
controlled-Z gate will be also accomplished. This can be clearly
seen
in Table \ref{t1}. 

\begin{figure}
\begin{center}
\includegraphics[width=7cm,angle=0]{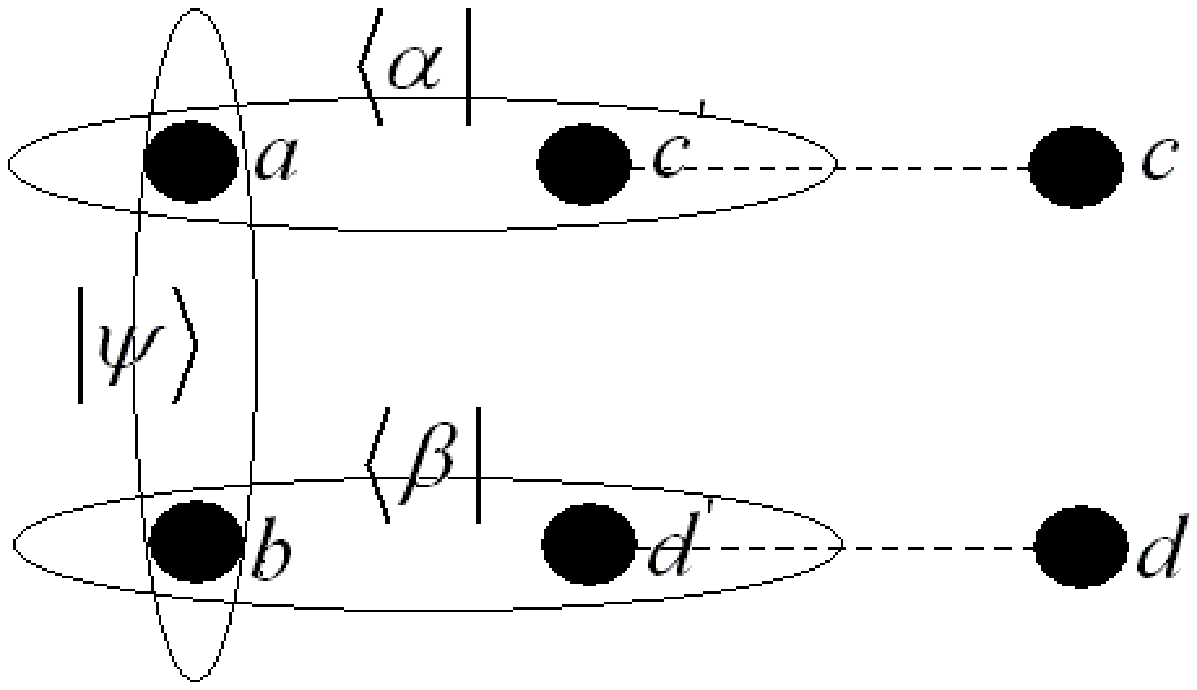}
\end{center}
\caption{ Without the crucial particles $e$ and $e\prime$, the
controlled-phase gate cannot be implemented.} \label{f3}
\end{figure}

We see the form of the entangled state  between $e$ and
$e^{\prime}$ is crucial in determining the measuring-basis for
implementing controlled-Z gate. Without $(e,e')$ in Fig.\ref{f3},
it is impossible to implement the controlled-Z gate.

\section{Construction of elementary quantum gates}
\label{s3}

\subsection{Generalized Controlled-Z Gates}
\label{ss1}

\begin{figure}
\begin{center}
\includegraphics[width=7cm,angle=0]{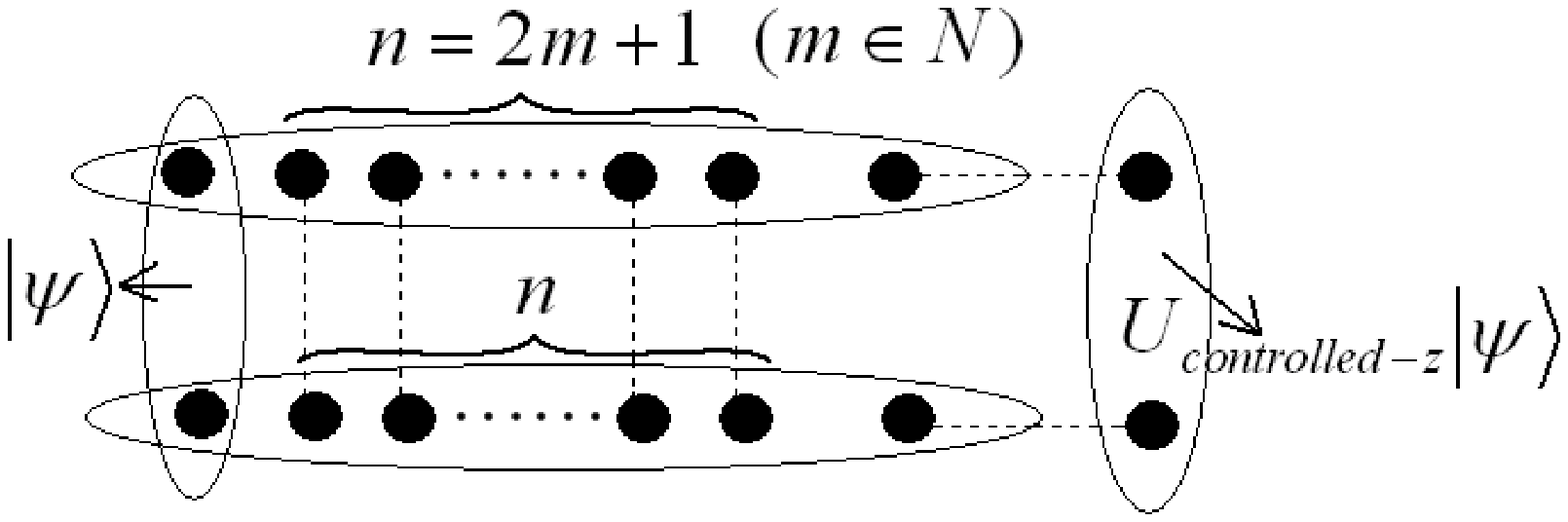}
\end{center}
\caption{ Implementation of the controlled-phase gate using $n$
pairs of singlet.} \label{f4a}
\end{figure}

The controlled-Z can be implemented with more number of singlets
and multi-qubit joint measurement. As in Fig.\ref{f4a}, the wave
function of $a$ and $b$ before joint measurement is
$|\psi\rangle=c_0|00\rangle+c_1|01\rangle+c_2|10\rangle+c_3|11\rangle$,
then the wave function of $c$ and $d$ after joint measurement is
$|\psi\prime\rangle=c_0|00\rangle+c_1|01\rangle+c_2|10\rangle+(-1)^nc_3|11\rangle$.
Therefore, the number of particles in a joint measurement must be
odd. This generalization is mathematically interesting, and it is
also of practical interest because in bulk quantum system this may
be the real case.

\begin{figure}
\begin{center}
\includegraphics[width=7cm,angle=0]{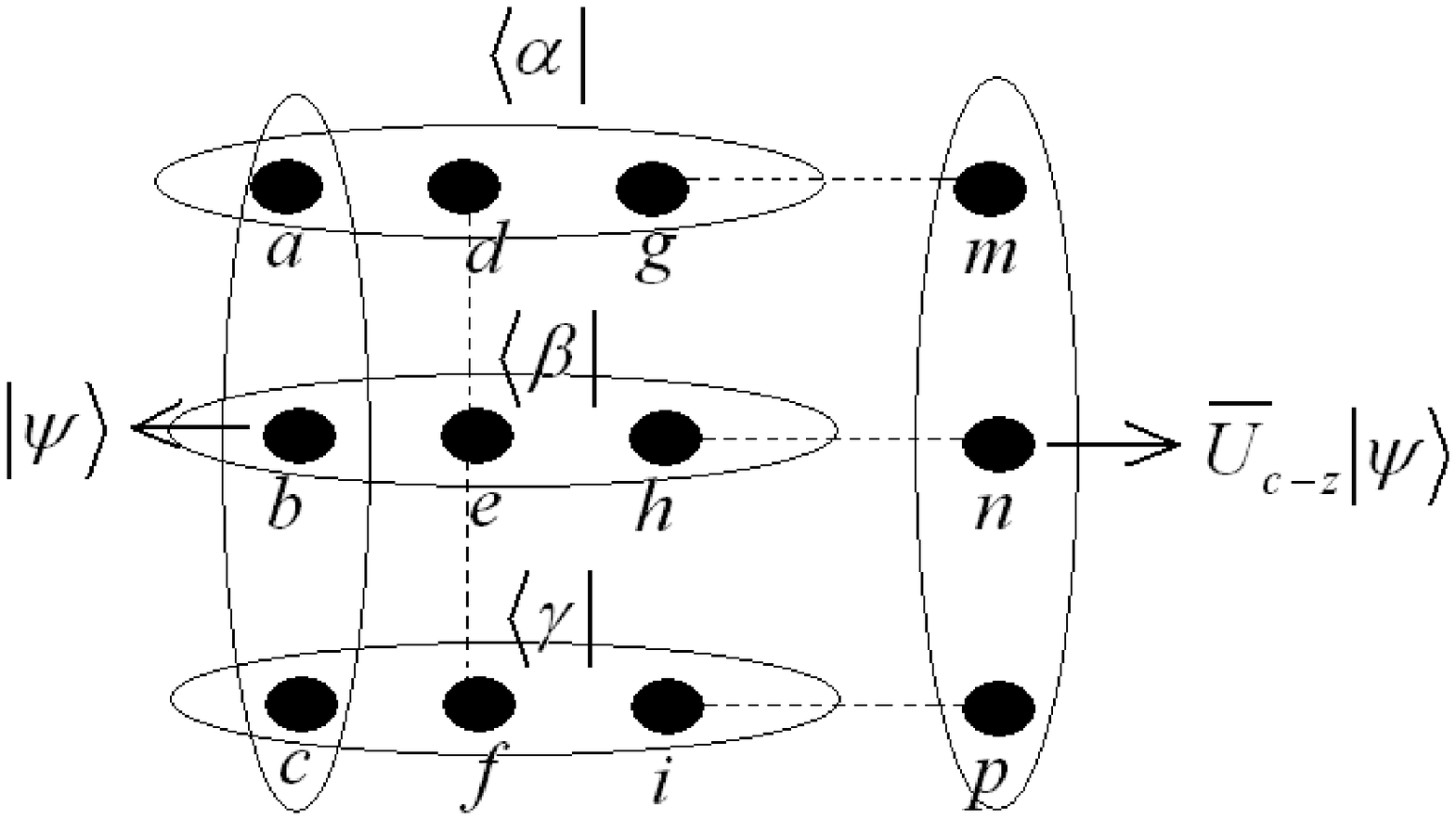}
\end{center}
\caption{ Implementation of triple-qubit controlled-phase gate by
making joint measurement on particle groups $(a,d,g)$ $(b,e,h)$
and $(c,f,i)$.} \label{f4b}
\end{figure}

As shown in Fig.\ref{f4b}, the entangled state of particles $d$,
$e$, $f$ is $\frac{1}{\sqrt{3}}(|000\rangle+|111\rangle)$ and the
entangled states of $g$ and $m$, $h$ and $n$, $i$ and $p$ are all
$\frac{1}{\sqrt{2}}(|00\rangle+|11\rangle)$. The state of
particles $a$, $b$ and $c$ is $|\psi\ket$. By making joint
measurement on particle groups $(a,d,g)$, $(b,e,h)$ and $(c, f,i)$
in the following measuring-basis,
\begin{eqnarray}
\{|\alpha\rangle\}&=&\{(\sigma_x)^i\otimes(\sigma_z)^j\otimes
I(|0_a+_d0_g\rangle\pm|1_a-_d1_g\rangle)\},
\label{e10}\\
\{|\beta\rangle\}&=&\{(\sigma_x)^i\otimes(\sigma_x)^j\otimes
I(|0_b0_e0_h\rangle\pm|1_b1_e1_h\rangle)\},
\label{e11}\\
\{|\gamma\rangle\}&=&\{(\sigma_x)^i\otimes(\sigma_z)^j\otimes
I(|0_c+_f0_i\rangle\pm|1_c-_f1_i\rangle)\},
\label{e12}
\end{eqnarray}
where $(i,j\in{\{0,1\}})$,  the following triple-qubit
controlled-Z gate
\begin{eqnarray}
\tilde{U}_{ctr-z}&=&|000\rangle\langle000|+|100\rangle\langle100|+|001\rangle\langle001|\nonumber\\
&&+|101\rangle\langle101|+|010\rangle\langle010|+|111\rangle\langle111|\nonumber\\
&&-|011\rangle\langle011|-|110\rangle\langle110|, \label{e13}
\end{eqnarray}
can be realized on the three resulting qubits $m$, $n$ and $p$.

\subsection{One-qubit Quantum Gates}

\subsubsection{Phase Gate}
\label{sss1}

\begin{figure}
\begin{center}
\includegraphics[width=7cm,angle=0]{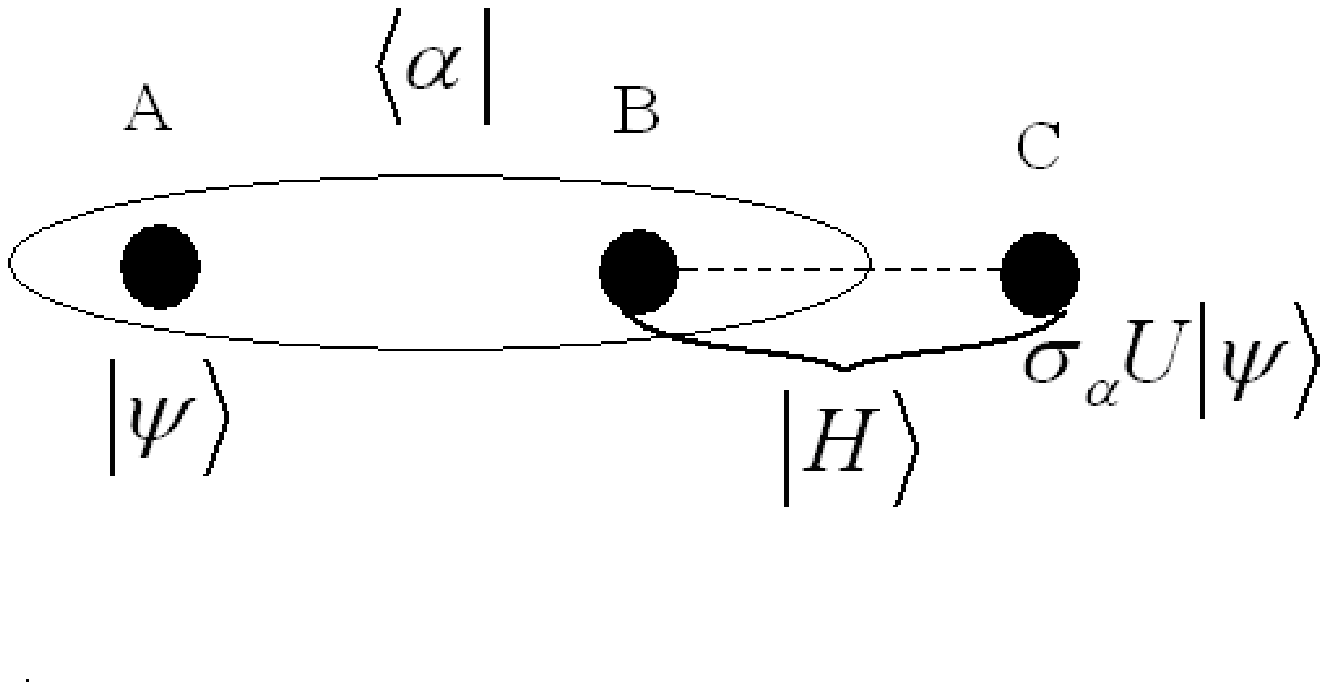}
\end{center}
\caption{ Implementation of a single-qubit unitary operation using
a pair of state in the $|H\ket$ entangled state.}\label{f1a}
\end{figure}

 By joint measurement, the simple one-qubit phase gate $U_{phase}=\left(%
\begin{array}{cc}
  1 & 0 \\
  0 & i \\
\end{array}%
\right)$ can be implemented by choosing the following complete
bases for joint measurement on qubit $A$ and $B$ as shown in
Fig.\ref{f1a}
$$\{|\alpha\rangle\}=\{\frac{|00\rangle+i|11\rangle}{\sqrt{2}},
\frac{|00\rangle-i|11\rangle}{\sqrt{2}},\frac{|01\rangle+i|10\rangle}{\sqrt{2}},
\frac{|01\rangle-i|10\rangle}{\sqrt{2}}\}$$ where  the entangled
state of particle $B$ and $C$ is
$\frac{1}{\sqrt{2}}(|00\rangle+|11\rangle)$. After the joint
measurement, a modified phase gate is applied to the resulting
particle $C$, and after an additional single qubit recovery
operation in the form of a Pauli matrix or the identity operator,
the phase gate is completed, and this is summarized in Table
\ref{t2}.
\begin{widetext}
\begin{center}
\begin{table}[tph]
\begin{center}
\caption{Construction of phase gate. The initial state is
$a|0\ket+b|1\ket$. }\label{t2}
\begin{tabular}{|c|c|c|}\hline
$|\alpha\ket$ &  Qubit $C$ state after measurement & Recovery operation\\
\hline $(|00\ket+i|11\ket)/\sqrt{2}$ & $a|0\ket-ib|1\ket$ & $\sigma_z$\\
\hline $(|00\ket-i|11\ket)/\sqrt{2}$ & $a|0\ket+ib|1\ket$ & $I$\\
\hline $(|01\ket+i|10\ket)/\sqrt{2}$ & $a|1\ket-ib|0\ket$ & $\sigma_z\sigma_x$\\
\hline $(|01\ket-i|10\ket)/\sqrt{2}$ & $a|1\ket+ib|0\ket$ & $\sigma_x$\\
\hline
\end{tabular}
\end{center}
\end{table}
\end{center}
\end{widetext}

\subsubsection{$\frac{\pi}{8}$ Gate.}
\label{sss2}

 The $\frac{\pi}{8}$ gate, which has the unitary matrix $$\left(%
\begin{array}{cc}
  1 & 0 \\
  0 & \frac{i\pi}{4} \\
\end{array}%
\right)$$ can be implemented by choosing the following
measuring-basis as
$\{|\alpha\rangle\}=\{\frac{|00\rangle+e^{\frac{-i\pi}{4}}|11\rangle}{\sqrt{2}}$,
$\frac{|00\rangle+e^{\frac{3i\pi}{4}}|11\rangle}{\sqrt{2}}$,
$\frac{|01\rangle+e^{\frac{-i\pi}{4}}|10\rangle}{\sqrt{2}},\frac{|01\rangle+e^{\frac{3i\pi}{4}}|10\rangle}{\sqrt{2}}\}$.
The corresponding states after the joint measurement and the extra
operations for the corresponding measured results are  given
explicitly  in Table \ref{t3}
\begin{widetext}
\begin{center}
\begin{table}[tph]
\begin{center}
\caption{Construction of $\pi/8$ gate. The initial state is
$a|0\ket+b|1\ket$}
\label{t3}
\begin{tabular}{|c|c|c|}\hline
$|\alpha\ket$ & Qubit $C$ state after measurement & Recovery Operation\\
\hline $(|00\ket+e^{-i\pi/4}|11\ket)/\sqrt{2}$ & $a|0\ket+e^{i\pi/4}b|1\ket$ & $I$\\
\hline $(|00\ket+e^{3i\pi/4}|11\ket)/\sqrt{2}$ & $a|0\ket+e^{-3i\pi/4}b|1\ket$ & $\sigma_z$\\
\hline $(|01\ket+e^{-i\pi/4}|10\ket)/\sqrt{2}$ & $a|1\ket+e^{i\pi/4}b|0\ket$ & $\sigma_x$\\
\hline $(|01\ket+e^{3i\pi/4}|10\ket)/\sqrt{2}$ & $a|1\ket+e^{-3i\pi/4}b|0\ket$ & $\sigma_z\sigma_x$\\
\hline
\end{tabular}
\end{center}
\end{table}
\end{center}
\end{widetext}

\subsection{Controlled-phase Gate}
\label{ss2}

Controlled-phase gate is important in constructing quantum
algorithms such as the Grover algorithm\cite{grover,long2} and in
in state initialization\cite{long3,knight}.  The usual
controlled-phase gate has been experimentally tested using NMR
technique\cite{Long}, and  cavity-QED realization of this gate has
been proposed\cite{Guo}. In the joint measurement scheme, by
combining the phase gate and controlled-Z gate, we can implement
controlled-phase gate
$$\left(%
\begin{array}{cccc}
  1 & 0 & 0 & 0 \\
  0 & 1 & 0 & 0 \\
  0 & 0 & 1 & 0 \\
  0 & 0 & 0 & i \\
\end{array}%
\right).$$  However one has to be careful in choosing the
measuring-basis. We first give an example. Suppose the
measuring-basis in Fig.\ref{f2a} are
\begin{eqnarray}
\{|\alpha\rangle\}&=&\{(\sigma_x)^i\otimes(\sigma_z)^j\otimes I(|0+0\rangle\pm{k}|1-1\rangle)\},\\
\label{e18}
\{|\beta\rangle\}&=&\{(\sigma_x)^i\otimes(\sigma_x)^j\otimes
I(|000\rangle\pm\tilde{k}|111\rangle)\},
\label{e19}
\end{eqnarray}
and the entangled state of $ee^{\prime}$ is
\begin{eqnarray}
\frac{1}{\sqrt{2}}(|0_e0_{e^{\prime}}\rangle+p|1_e1_{e^{\prime}}\rangle),
\label{e20}
\end{eqnarray}
the entangled state of $cc^{\prime}$ is
\begin{eqnarray}
\frac{1}{\sqrt{2}}(|0_c0_{c^{\prime}}\rangle+m|1_c1_{c^{\prime}}\rangle),
\label{e21}
\end{eqnarray}
and the entangled state of $dd^{\prime}$ is
\begin{eqnarray}
\frac{1}{\sqrt{2}}(|0_d0_{d^{\prime}}\rangle+n|1_d1_{d^{\prime}}\rangle).
\label{e22}
\end{eqnarray}
The normalization condition requires that
$kk^{*}=\tilde{k}\tilde{k}^{*}=pp^{*}=mm^{*}=nn^{*}=1$.  Then the
joint measurement will turn the state of  $c$ and $d$ into
\begin{eqnarray}
U=\left(%
\begin{array}{cccc}
  1 & 0 & 0 & 0 \\
  0 & np\tilde{k}^{*} & 0 & 0 \\
  0 & 0 & mk^{*} & 0 \\
  0 & 0 & 0 & -mnpk^{*}\tilde{k}^{*} \\
\end{array}%
\right), \label{e23}
\end{eqnarray}
which cannot fulfil the controlled-phase gate obviously  and when
all the parameters reduce to 1 then $U$ turns to be controlled-Z
gate. In this case, if we still use the entangled states for the
pairs of qubits as given
 in Eqns. (\ref{e20},\ref{e21},\ref{e22}),
 the appropriate measuring-basis are
\begin{eqnarray}
\{|\alpha\rangle\}&=&\{(\sigma_x)^j\otimes(\sigma_z)^k\otimes
I(|0\rangle|+\ket|0\rangle\pm|1\rangle(\frac{|0\rangle-i|1\rangle}{\sqrt{2}})|1\rangle)\},\nonumber\\
\label{e24}
\{|\beta\rangle\}&=&\{(\sigma_x)^j\otimes(\sigma_x)^k\otimes
I(|000\rangle\pm|111\rangle)\}. \label{e25}
\end{eqnarray}

For simplicity we take the entangled states between the
$cc^{\prime}$, $ee^{\prime}$, $dd^{\prime}$ pairs as
$\frac{1}{\sqrt{2}}(|00\rangle+|11\rangle)
    (j,k\in\{0,1\})$.
It is interesting to note that although the inner two states
mentioned in measuring-basis
    $\{|\alpha\ket\}$, i.e. $|+\rangle$ and $\frac{|0\rangle-i|1\rangle}{\sqrt{2}}$
are not orthogonal each other, the bases in $\{|\alpha\rangle\}$
are themselves orthogonal. It is because the total Hilbert space
is eight-dimensional and even though
$\frac{\langle0|+\langle1|}{\sqrt{2}}|\frac{|0\rangle-i|1\rangle}{\sqrt{2}}\neq0$,
the inner two parts of $\{|\alpha\ket\}$ ($e.g.$
$|0\rangle|+\rangle|0\rangle$ and
$|1\rangle\frac{|0\rangle-i|1\rangle}{\sqrt{2}}|1\rangle$) are
orthogonal. The only request here is that
$\frac{\langle0|+\langle1|}{\sqrt{2}}|(\sigma_z\frac{|0\rangle
+|1\rangle}{\sqrt{2}})$=$\frac{\langle0|+i\langle1|}{\sqrt{2}}|(\sigma_z\frac{|0\rangle-i|1\rangle}{\sqrt{2}})$=0.
In particular, the phase $i$ comes from
 that in $\{|\alpha\rangle\}$ basis set.

\begin{widetext}
\begin{center}
\begin{table}[tph]
\begin{center}
\caption{Construction of controlled-phase gate. The rows are
$(j,k,\pm)$ in $\{|\alpha\ket\}$ and the columns are $(j,k,\pm)$
in $\{|\beta\ket\}$. The quantity is the appropriate operation to
be done after the joint measurement}\label{t4}
\begin{tabular}{|c|c|c|c|c|}\hline
$ $ & $(0,0,+)$ & $(0,0,-)$ & $(0,1,+)$ & $(0,1,-)$\\
\hline $(0,0,+)$ & $I$ & $\sigma_z\otimes I$ & $I\otimes\sigma_z$
&
$\sigma_z\otimes\sigma_z$\\
\hline $(0,0,-)$ & $I\otimes\sigma_z$ & $\sigma_z\otimes\sigma_z$
&
$I$ & $\sigma_z\otimes I$\\
\hline $(0,1,+)$ & $U_{cz}(\sigma_zU_p\otimes I)$ &
$U_{cz}(U_p\otimes I)$
& $U_{cz}(\sigma_zU_p\otimes\sigma_z)$ & $U_{cz}(U_p\otimes\sigma_z)$\\
\hline $(0,1,-)$ & $U_{cz}(\sigma_z U_p\otimes\sigma_z)$ &
$U_{cz}(U_p\otimes\sigma_z)$ & $U_{cz}(\sigma_z U_p\otimes I)$ &
$U_{cz}(U_p\otimes I)$\\
\hline $(1,0,+)$ & $U_{cz}(\sigma_zU_p\otimes\sigma_x)$ &
$U_{cz}(U_p\otimes\sigma_x)$ &
$U_{cz}(\sigma_zU_p\otimes\sigma_z\sigma_x)$ &
$U_{cz}(U_p\otimes\sigma_z\sigma_x)$\\
\hline $(1,0,-)$ & $U_{cz}(\sigma_zU_p\otimes\sigma_x\sigma_z)$ &
$U_{cz}(U_p\otimes\sigma_x\sigma_z)$ &
$U_{cz}(\sigma_zU_p\otimes\sigma_x)$ & $U_{cz}(U_p\otimes\sigma_x)$\\
\hline $(1,1,+)$ & $I\otimes\sigma_x$ & $\sigma_z\otimes\sigma_x$
&
$I\otimes\sigma_z\sigma_x$ & $\sigma_z\otimes\sigma_z\sigma_x$\\
\hline $(1,1,-)$ & $I\otimes\sigma_x\sigma_z$ &
$\sigma_z\otimes\sigma_x\sigma_z$ & $I\otimes\sigma_x$ &
$\sigma_z\otimes\sigma_x$\\
\hline\hline

$ $ & $(1,0,+)$ & $(1,0,-)$ & $(1,1,+)$ & $(1,1,-)$\\
\hline $(0,0,+)$ & $U_{cz}(\sigma_x\otimes\sigma_zU_p)$ &
$U_{cz}(\sigma_x\sigma_z\otimes\sigma_zU_p)$ &
$U_{cz}(\sigma_x\otimes U_p)$ &
$U_{cz}(\sigma_z\sigma_x\otimes U_p)$\\
\hline $(0,0,-)$ & $U_{cz}(\sigma_x\otimes U_p)$ &
$U_{cz}(\sigma_x\sigma_z\otimes U_p)$ &
$U_{cz}(\sigma_x\otimes U_p\sigma_z)$ & $U_{cz}(\sigma_z\sigma_x\otimes U_p\sigma_z)$\\
\hline $(0,1,+)$ & $U_p\sigma_x\otimes U_p$ & $\sigma_z
U_p\sigma_x\otimes U_p$ & $U_p\sigma_x\otimes U_p\sigma_z$ &
$\sigma_z U_p\sigma_x\otimes\sigma_z U_p$\\
\hline $(0,1,-)$ & $U_p\sigma_x\otimes\sigma_zU_p$ &
$\sigma_zU_p\sigma_x\otimes\sigma_zU_p$ & $U_p\sigma_x\otimes U_p$
&
$\sigma_zU_p\sigma_x\otimes U_p$\\
\hline $(1,0,+)$ & $U_p\sigma_x\otimes U_p\sigma_x$ &
$\sigma_zU_p\sigma_x\otimes U_p\sigma_x$ &
$U_p\sigma_x\otimes\sigma_z U_p\sigma_x$ &
$\sigma_z U_p\sigma_x\otimes\sigma_z U_p\sigma_x$\\
\hline $(1,0,-)$ & $U_p\sigma_x\otimes\sigma_z U_p\sigma_x$ &
$\sigma_z U_p\sigma_x\otimes\sigma_z U_p\sigma_x$ &
$U_p\sigma_x\otimes
U_p\sigma_x$ & $\sigma_zU_p\sigma_x\otimes U_p\sigma_x$\\
\hline $(1,1,+)$ & $U_{cz}(\sigma_x\otimes\sigma_z U_p\sigma_x)$ &
$U_{cz}(\sigma_x\sigma_z\otimes\sigma_zU_p\sigma_x)$ &
$U_{cz}(\sigma_x\otimes U_p\sigma_x)$ &
$U_{cz}(\sigma_z\sigma_x\otimes U_p\sigma_x)$\\
\hline $(1,1,-)$ & $U_{cz}(\sigma_x\otimes U_p\sigma_x)$ &
$U_{cz}(\sigma_x\sigma_z\otimes U_p\sigma_x)$ &
$U_{cz}(\sigma_x\otimes\sigma_z U_p\sigma_x)$ & $U_{cz}(\sigma_z\sigma_x\otimes\sigma_z U_p\sigma_x)$\\
\hline

\end{tabular}
\end{center}
\end{table}
\end{center}
\end{widetext}

\begin{figure}
\begin{center}
\includegraphics[width=7cm,angle=90]{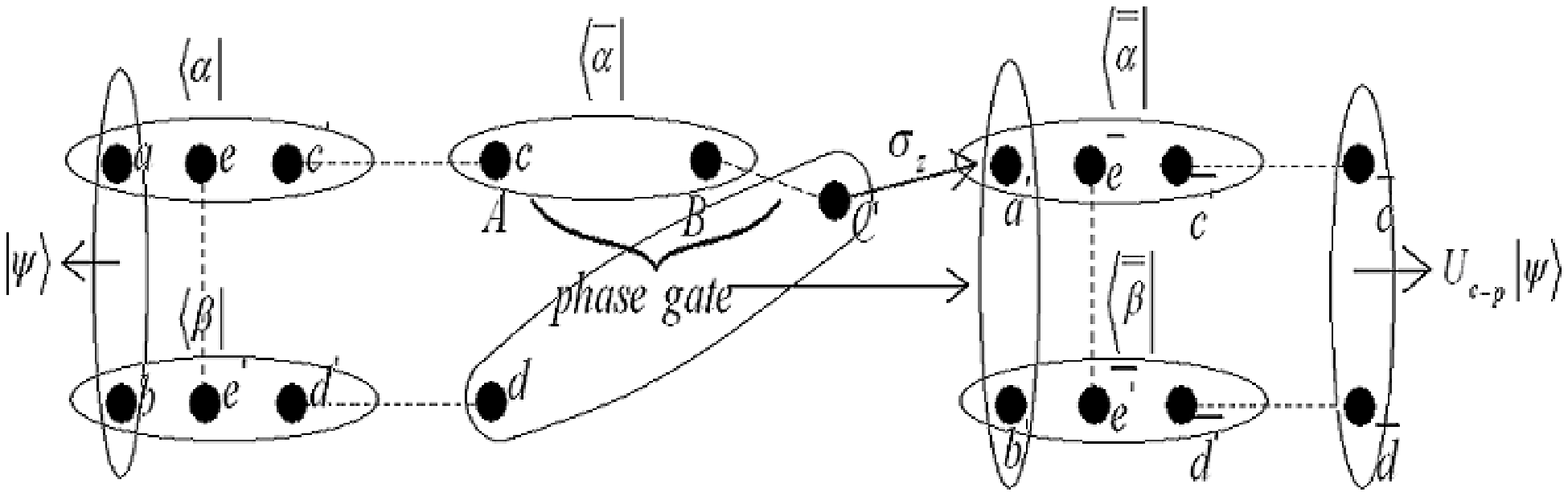}
\end{center}
\caption{ An additional operation of $U_{cz}\sigma_zU_p\otimes I$
on particles $c$ and $d$ for implementing controlled-phase gate if
the result of the joint measurement is
$|\alpha\rangle=|0\rangle_a(\frac{|0\rangle+|1\rangle}{\sqrt{2}})_e|0\rangle_{c^{\prime}}
+|1\rangle_a(\frac{|0\rangle-i|1\rangle}{\sqrt{2}})_e|1\rangle_{c^{\prime}}$
and
$|\beta\rangle=|0\rangle_b|1\rangle_{e^{\prime}}|0\rangle_{d^{\prime}}
+|1\rangle_b|0\rangle_{e^{\prime}}|1\rangle_{d^{\prime}}$. It
should be noted that $\{|\bar{\alpha}\rangle\}$ is the same as in
Table \ref{t2} and
$\{|\bar{\bar{\alpha}}\rangle\}$=$\{(\sigma_x)^i\otimes(\sigma_z)^j\otimes
I(|0+0\rangle\pm|1-1\rangle)\}$;
$\{|\bar{\bar{\beta}}\rangle\}$=$\{(\sigma_x)^i\otimes(\sigma_x)^j\otimes
I(|000\rangle\pm|111\rangle)\}.$ Besides, all the entangled pairs
in this figure are in the form
$\frac{|00\rangle+|11\rangle}{\sqrt{2}}.$} \label{f5}
\end{figure}

The details of the construction of the controlled-phase gate are
given in  Table \ref{t4}. For example, if the result of the joint
measurement is
$|\alpha\rangle=|0\rangle_a(\frac{|0\rangle+|1\rangle}{\sqrt{2}})_e|0\rangle_{c^{\prime}}
+|1\rangle_a(\frac{|0\rangle-i|1\rangle}{\sqrt{2}})_e|1\rangle_{c^{\prime}}$
and
$|\beta\rangle=|0\rangle_b|1\rangle_{e^{\prime}}|0\rangle_{d^{\prime}}
+|1\rangle_b|0\rangle_{e^{\prime}}|1\rangle_{d^{\prime}}$ then the
controlled-phase gate can be implemented by an additional
operation of $U_{cz}(\sigma_zU_p\otimes I) $ as shown
Fig.\ref{f5}. It should be noted that $
\{\sigma_z,\;\sigma_x\}$=$[\sigma_z,\;U_p]$=0, $U_p$ is phase gate
and $U_{cz}$ is controlled-Z gate.

\subsection{Controlled-NOT Gate}
\label{ss3}

Controlled-NOT gate is important in quantum computation. Great
interests are attached to this gate. Recent experimental
demonstration of this controlled-NOT gate has been reported in
\cite{Brien,Zeilinger}. While by joint measurement, the
implementation of the controlled-NOT gate (C-NOT) is not
straight-forward as one might have expected. After some tedious
calculation, we find that to construct C-NOT gate, one has to make
sure that the two highlighted parts  in the following equation
should be different.  The swapping between
$|0_{d^{\prime}}\rangle$ and $|1_{d^{\prime}}\rangle$ in the
second highlighted part leads to the switching in C-NOT gate. The
minus sign in that term can be cancelled in the calculation( see
also Figs.\ref{f2a},\ref{f2b} for illustration )
\begin{widetext}
\begin{center}
\begin{eqnarray}
\langle\alpha|\Phi\rangle&=&(c_{00}|0_b\rangle+c_{01}|1_b\rangle)\otimes\frac{1}{2}
(|0_{e^{\prime}}\rangle+|1_{e^{\prime}}\rangle)\otimes\frac{1}{\sqrt{2}}|0_c\rangle
\otimes\mathbf{\frac{1}{2}(|0_d0_{d^{\prime}}\rangle+|1_d1_{d^{\prime}}\rangle)}\nonumber\\
&&+(c_{10}|0_b\rangle+c_{11}|1_b\rangle)\otimes\frac{1}{2}(|0_{e^{\prime}}\rangle-|1_{e^{\prime}}\rangle)
\otimes\frac{1}{\sqrt{2}}|1_c\rangle\otimes\mathbf{\frac{1}{2}(-|0_d1_{d^{\prime}}\rangle+|1_d0_{d^{\prime}}\rangle)}.
\label{e26}
\end{eqnarray}
\end{center}
\end{widetext}

\begin{figure}
\begin{center}
\includegraphics[width=7cm,angle=0]{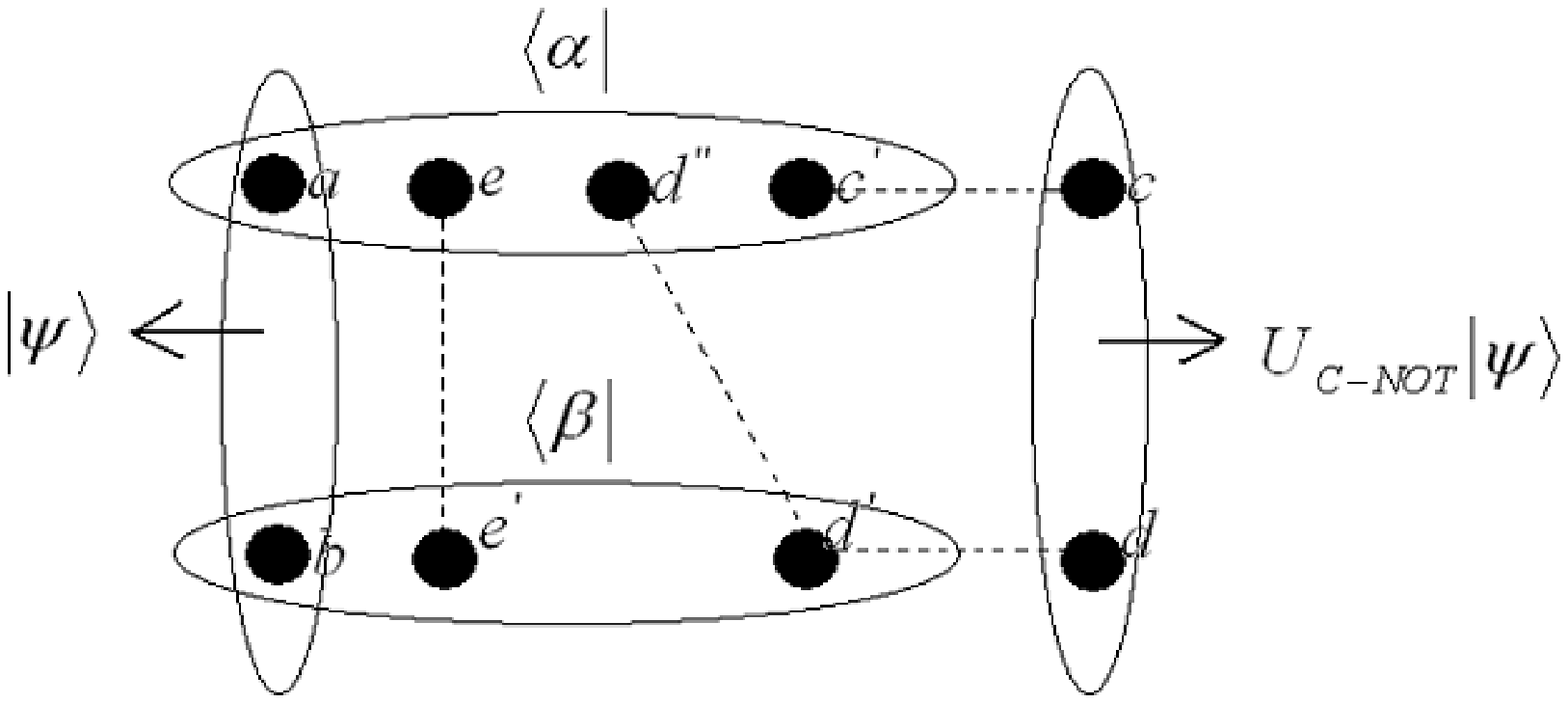}
\end{center}
\caption{ Implementation of C-NOT gate by joint measurement using
singlets and three-particle entanglement.} \label{f6}
\end{figure}

We note that this difference is crucial in the process of
implementing the C-NOT gate. We can fulfill this by adding an
extra particle $d^{\prime\prime}$ (see Fig.\ref{f6}) in the first
group $\{|\alpha\rangle\}$ for joint measurement. We give some
necessary ingredients below.

The entangled state of $ee^{\prime}$ is
\begin{eqnarray}
\frac{1}{\sqrt{2}}(|0_e0_{e^{\prime}}\rangle+|1_e1_{e^{\prime}}\rangle).
\label{e27}
\end{eqnarray}

The entangled state of $cc^{\prime}$ is
\begin{eqnarray}
\frac{1}{\sqrt{2}}(|0_c0_{c^{\prime}}\rangle+|1_c1_{c^{\prime}}\rangle).
\label{e28}
\end{eqnarray}

The entangled state of $dd^{\prime}d^{\prime\prime}$ is
\begin{eqnarray}
\frac{1}{2}(|0_d0_{d^{\prime}}0_{d^{\prime\prime}}\rangle+|1_d1_{d^{\prime}}0_{d^{\prime\prime}}
\rangle+|1_d0_{d^{\prime}}1_{d^{\prime\prime}}\rangle-|0_d1_{d^{\prime}}1_{d^{\prime\prime}}\rangle).
\label{e29}
\end{eqnarray}

The measuring-basis $\{|\alpha\ket\}$  and  $\{|\beta\ket\}$ are
\begin{eqnarray}
\{|\alpha\rangle\}&=&\{(\sigma_x)^i\otimes(\sigma_z)^j\otimes(\sigma_x)^k\otimes I\nonumber\\
&&(|0_a+_e0_{c^{\prime}}0_{d^{\prime\prime}}\rangle\pm|1_a-_e1_{c^{\prime}}1_{d^{\prime\prime}}\rangle)\},\nonumber\\
\label{e30}
\{|\beta\rangle\}&=&\{(\sigma_x)^i\otimes(\sigma_x)^j\otimes I\nonumber\\
&&(|0_b0_{e^{\prime}}0_{d^{\prime}}\rangle\pm|1_b1_{e^{\prime}}1_{d^{\prime}}\rangle)\},
\label{e31}
\end{eqnarray}
where
$(i,j,k\in\{0,1\};|\pm\rangle=\frac{1}{\sqrt{2}}(|0\rangle\pm|1\rangle))$.
We choose the measuring-basis for particles $e$ and $e\prime$ as
$|\pm_e\rangle$ and $|0_{e\prime}\rangle(|1_{e\prime}\rangle)$ to
avoid the loss of information which was mentioned earlier in
section \ref{s2}. Of course after the joint measurement, some
extra operations with the form $\sigma_\alpha$ $ (\alpha=1,2,3,4;
\sigma_4= I)$ need to be applied to the wave function of particles
$c$ and $d$ as in other operations as well. The details of
construction of C-NOT gate are given in Table \ref{t5}

\begin{widetext}
\begin{center}
\begin{table}[tph]
\begin{center}
\caption{Construction of C-NOT gate. The rows are $(i,j,k,\pm)$ in
$\{|\alpha\ket\}$ and the columns are $(i,j,\pm)$ in
$\{|\beta\ket\}$. The quantity is the appropriate operation to be
done after the joint measurement}\label{t5}
\begin{tabular}{|c|c|c|c|c|c|c|c|c|}\hline
$ $ & $(0,0,0,+)$ & $(0,0,0,-)$ & $(0,0,1,+)$ & $(0,0,1,-)$ & $(0,1,0,+)$ & $(0,1,0,-)$ & $(0,1,1,+)$ & $(0,1,1,-)$\\
\hline $(0,0,+)$ & $I$ & $\sigma_z\otimes I$ & $\sigma_x\otimes I$
&
$\sigma_z\sigma_x\otimes I$ & $I\otimes\sigma_z$ & $\sigma_z\otimes\sigma_z$ & $\sigma_x\otimes\sigma_z$ & $\sigma_z\sigma_x\otimes\sigma_z$\\
\hline $(0,0,-)$ & $I\otimes\sigma_z$ & $\sigma_z\otimes\sigma_z$
&
$\sigma_x\otimes\sigma_z$ & $\sigma_z\sigma_x\otimes\sigma_z$ & $I$ & $\sigma_z\otimes I$ & $\sigma_x\otimes I$ & $\sigma_z\sigma_x\otimes I$\\
\hline $(0,1,+)$ & $\sigma_z\otimes I$ & $I$
& $\sigma_z\sigma_x\otimes I$ & $\sigma_x\otimes I$ & $\sigma_z\otimes\sigma_z$ & $I\otimes\sigma_z$ & $\sigma_z\sigma_x\otimes\sigma_z$ & $\sigma_x\otimes\sigma_z$\\
\hline $(0,1,-)$ & $\sigma_z\otimes\sigma_z$ & $I\otimes\sigma_z$
& $\sigma_z\sigma_x\otimes\sigma_z$ &
$\sigma_x\otimes\sigma_z$ & $\sigma_z\otimes I$ & $I$ & $\sigma_z\sigma_x\otimes I$ & $\sigma_x\otimes I$\\
\hline $(1,0,+)$ & $I\otimes\sigma_x$ & $\sigma_z\otimes\sigma_x$
& $\sigma_x\otimes\sigma_x$ &
$\sigma_z\sigma_x\otimes\sigma_x$ & $I\otimes\sigma_z\sigma_x$ & $\sigma_z\otimes\sigma_z\sigma_x$ & $\sigma_x\otimes\sigma_z\sigma_x$ & $\sigma_z\sigma_x\otimes\sigma_z\sigma_x$\\
\hline $(1,0,-)$ & $I\otimes\sigma_z\sigma_x$ &
$\sigma_z\otimes\sigma_z\sigma_x$ &
$\sigma_x\otimes\sigma_z\sigma_x$ & $\sigma_z\sigma_x\otimes\sigma_z\sigma_x$ & $I\otimes\sigma_x$ & $\sigma_z\otimes\sigma_x$ & $\sigma_x\otimes\sigma_x$ & $\sigma_z\sigma_x\otimes\sigma_x$\\
\hline $(1,1,+)$ & $\sigma_z\otimes\sigma_x$ & $I\otimes\sigma_x$
&
$\sigma_z\sigma_x\otimes\sigma_x$ & $\sigma_x\otimes\sigma_x$ & $\sigma_z\otimes\sigma_z\sigma_x$ & $I\otimes\sigma_z\sigma_x$ & $\sigma_z\sigma_x\otimes\sigma_z\sigma_x$ & $\sigma_x\otimes\sigma_z\sigma_x$\\
\hline $(1,1,-)$ & $\sigma_z\otimes\sigma_z\sigma_x$ &
$I\otimes\sigma_z\sigma_x$ &
$\sigma_z\sigma_x\otimes\sigma_z\sigma_x$ &
$\sigma_x\otimes\sigma_z\sigma_x$ & $\sigma_z\otimes\sigma_x$ & $I\otimes\sigma_x$ & $\sigma_z\sigma_x\otimes\sigma_x$ & $\sigma_x\otimes\sigma_x$\\
\hline\hline
$ $ & $(1,0,0,+)$ & $(1,0,0,-)$ & $(1,0,1,+)$ & $(1,0,1,-)$ & $(1,1,0,+)$ & $(1,1,0,-)$ & $(1,1,1,+)$ & $(1,1,1,-)$\\
\hline $(0,0,+)$ & $\sigma_x\otimes\sigma_x$ &
$\sigma_z\sigma_x\otimes\sigma_x$ & $I\otimes\sigma_x$ &
$\sigma_z\otimes\sigma_x$ & $\sigma_x\otimes\sigma_z\sigma_x$ & $\sigma_z\sigma_x\otimes\sigma_z\sigma_x$ & $I\otimes\sigma_z\sigma_x$ & $\sigma_z\otimes\sigma_z\sigma_x$\\
\hline $(0,0,-)$ & $\sigma_x\otimes\sigma_z\sigma_x$ &
$\sigma_z\sigma_x\otimes\sigma_z\sigma_x$ &
$I\otimes\sigma_z\sigma_x$ & $\sigma_z\otimes\sigma_z\sigma_x$ & $\sigma_x\otimes\sigma_x$ & $\sigma_z\sigma_x\otimes\sigma_x$ & $I\otimes\sigma_x$ & $\sigma_z\otimes\sigma_x$\\
\hline $(0,1,+)$ & $\sigma_z\sigma_x\otimes\sigma_x$ &
$\sigma_x\otimes\sigma_x$
& $\sigma_z\otimes\sigma_x$ & $I\otimes\sigma_x$ & $\sigma_z\sigma_x\otimes\sigma_z\sigma_x$ & $\sigma_x\otimes\sigma_z\sigma_x$ & $\sigma_z\otimes\sigma_z\sigma_x$ & $I\otimes\sigma_z\sigma_x$\\
\hline $(0,1,-)$ & $\sigma_z\sigma_x\otimes\sigma_z\sigma_x$ &
$\sigma_x\otimes\sigma_z\sigma_x$ &
$\sigma_z\otimes\sigma_z\sigma_x$ &
$I\otimes\sigma_z\sigma_x$ & $\sigma_z\sigma_x\otimes\sigma_x$ & $\sigma_x\otimes\sigma_x$ & $\sigma_z\otimes\sigma_x$ & $I\otimes\sigma_x$\\
\hline $(1,0,+)$ & $\sigma_x\otimes I$ & $\sigma_z\sigma_x\otimes
I$ & $I$ &
$\sigma_z\otimes I$ & $\sigma_x\otimes\sigma_z$ & $\sigma_z\sigma_x\otimes\sigma_z$ & $I\otimes\sigma_z$ & $\sigma_z\otimes\sigma_z$\\
\hline $(1,0,-)$ & $\sigma_x\otimes\sigma_z$ &
$\sigma_z\sigma_x\otimes\sigma_z$ &
$I\otimes\sigma_z$ & $\sigma_z\otimes\sigma_z$ & $\sigma_x\otimes I$ & $\sigma_z\sigma_x\otimes I$ & $I$ & $\sigma_z\otimes I$\\
\hline $(1,1,+)$ & $\sigma_z\sigma_x\otimes I$ & $\sigma_x\otimes
I$ &
$\sigma_z\otimes I$ & $I$ & $\sigma_z\sigma_x\otimes\sigma_z$ & $\sigma_x\otimes\sigma_z$ & $\sigma_z\otimes\sigma_z$ & $I\otimes\sigma_z$\\
\hline $(1,1,-)$ & $\sigma_z\sigma_x\otimes\sigma_z$ &
$\sigma_x\otimes\sigma_z$ & $\sigma_z\otimes\sigma_z$ &
$I\otimes\sigma_z$ & $\sigma_z\sigma_x\otimes I$ & $\sigma_x\otimes I$ & $\sigma_z\otimes I$ & $I$\\
\hline

\end{tabular}
\end{center}
\end{table}
\end{center}
\end{widetext}

\subsection{Swap Gate}
\label{ss4}

\begin{figure}
\begin{center}
\includegraphics[width=7cm,angle=0]{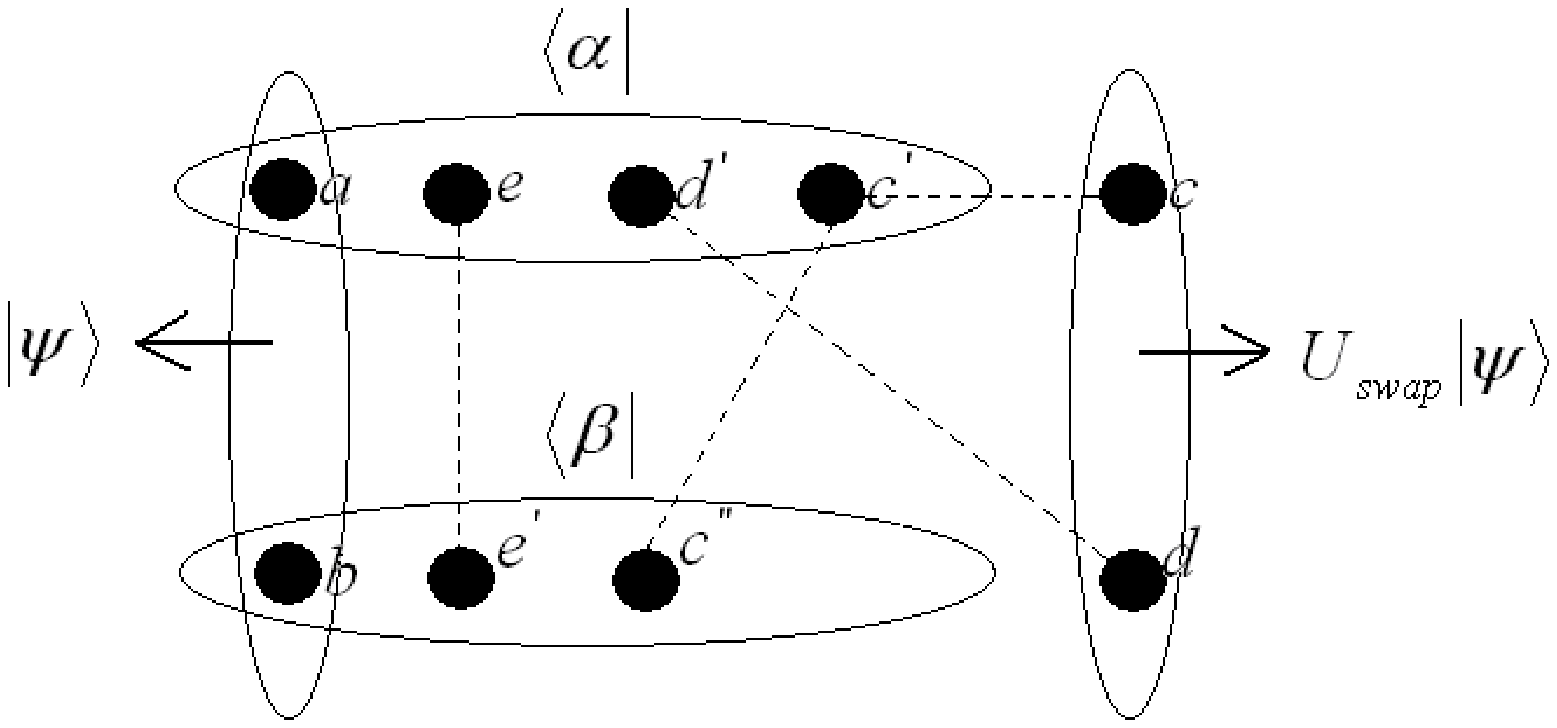}
\end{center}
\caption{ Implementation of swap gate by joint measurement using
singlets and GHZ-like state. Note that the pattern of entangled
states in this process is the same as that in Fig. \ref{f6}.}
\label{f7}
\end{figure}

The swap gate matrix is
$$\left(%
\begin{array}{cccc}
  1 & 0 & 0 & 0 \\
  0 & 0 & 1 & 0 \\
  0 & 1 & 0 & 0 \\
  0 & 0 & 0 & 1 \\
\end{array}%
\right).$$  In the joint measurement scheme, swap gate can be
implemented similarly to the C-NOT gate. Attentions have to be
paid to the four highlighted parts in the following equation,
which is also illustrated in Figs.\ref{f2a},\ref{f2b} and
\ref{f7},
\begin{eqnarray}
\langle\alpha|\Phi\rangle&&=(c_{00}|0_b\rangle+c_{01}|1_b\rangle)\otimes\frac{1}{2}(|0_{e^{\prime}}\rangle+|1_{e^{\prime}}\rangle)\otimes\nonumber\\
&&\mathbf{\frac{1}{2}(|0_c0_{c^{\prime\prime}}\rangle+|1_c1_{c^{\prime\prime}}\rangle)}\otimes\mathbf{\frac{1}{2}|0_d\rangle}+\nonumber\\
&&(c_{10}|0_b\rangle+c_{11}|1_b\rangle)\otimes\frac{1}{2}(|0_{e^{\prime}}\rangle-|1_{e^{\prime}}\rangle)\otimes\nonumber\\
&&\mathbf{\frac{1}{2}(|0_c0_{c^{\prime\prime}}\rangle-|1_c1_{c^{\prime\prime}}\rangle)}\otimes\mathbf{\frac{1}{2}|1_d\rangle}.\nonumber\\
\label{e32}
\end{eqnarray}
They should not be the same.
 The collapse of the quantum state of
particle $d$ to $|0_d\rangle$ and $|1_d\rangle$ leads to the
fulfillment of swap gate. In order to make this difference, we
should measure $d$ in the first group $\{|\alpha\rangle\}$. The
minus sign in the highlighted term
$\frac{1}{2}(|0_c0_{c\prime\prime}\rangle-|1_c1_{c\prime\prime}\rangle)$
can be cancelled in the calculation.  The scheme is illustrated in
Fig.\ref{f7}. Explicitly the state wave functions of the pairs and
the corresponding measuring-basis sets are given respectively
below.

The entangled state of $ee^{\prime}$ is
\begin{eqnarray}
\frac{1}{\sqrt{2}}(|0_e0_{e^{\prime}}\rangle+|1_e1_{e^{\prime}}\rangle),
\label{e33}
\end{eqnarray}

The entangled state of $dd^{\prime}$ is
\begin{eqnarray}
\frac{1}{\sqrt{2}}(|0_d0_{d^{\prime}}\rangle+|1_d1_{d^{\prime}}\rangle),
\label{e34}
\end{eqnarray}

The  GHZ-like entangled state of $cc^{\prime}c^{\prime\prime}$ is
\begin{eqnarray}
\frac{1}{2}(|0_c0_{c^{\prime}}0_{c^{\prime\prime}}\rangle+|1_c0_{c^{\prime}}1_{c^{\prime\prime}}\rangle+|0_c1_{c^{\prime}}0_{c^{\prime\prime}}\rangle-|1_c1_{c^{\prime}}1_{c^{\prime\prime}}\rangle)=\frac{1}{\sqrt{2}}(\frac{|0_c0_{c\prime\prime}\rangle+|1_c1_{c\prime\prime}\rangle}{\sqrt{2}}|0_{c\prime}\rangle+\frac{|0_c0_{c\prime\prime}\rangle-|1_c1_{c\prime\prime}\rangle}{\sqrt{2}}|1_{c\prime}\rangle),
\label{e35}
\end{eqnarray}

We can add a Hadamard operation on $c^{\prime}$ to change this
entangled state to the term
$\frac{1}{\sqrt{2}}(|0_c0_{c^{\prime}}0_{c^{\prime\prime}}\rangle+|1_c1_{c^{\prime}}1_{c^{\prime\prime}}\rangle)$.

The measuring-basis sets are
\begin{eqnarray}
\{|\alpha\rangle\}&=&\{(\sigma_x)^i\otimes(\sigma_z)^j\otimes I\otimes(\sigma_x)^k\nonumber\\
&&(|0_a+_e0_{c^{\prime}}0_{d^{\prime}}\rangle\pm|1_a-_e1_{c^{\prime}}1_{d^{\prime}}\rangle)\},
\label{e36}\\
\{|\beta\rangle\}&=&\{(\sigma_x)^i\otimes(\sigma_x)^j\otimes I\nonumber\\
&&(|0_b0_{e^{\prime}}0_{c^{\prime\prime}}\rangle\pm|1_b1_{e^{\prime}}
1_{c^{\prime\prime}}\rangle)\}, \label{e37}
\end{eqnarray}
where $(i,j,k\in\{0,1\})$. It is worth pointing that particles and
$c^{\prime\prime}$ $d^{\prime}$ are crucial in implementing this
quantum gate. The pattern of entangled states for constructing
swap gate is just the same as that for constructing the C-NOT
gate. The difference is at the distinct ways for the joint
measurement and the different three-particle entangled states
respectively. The details of construction of swap gate are given
in Table \ref{t6}.

\begin{widetext}
\begin{center}
\begin{table}[tph]
\begin{center}
\caption{Construction of swap gate. The rows are $(i,j,k,\pm)$ in
$\{|\alpha\ket\}$ and the columns are $(i,j,\pm)$ in
$\{|\beta\ket\}$. The quantity is the appropriate operation to be
done after the joint measurement}\label{t6}
\begin{tabular}{|c|c|c|c|c|c|c|c|c|}\hline
$ $ & $(0,0,0,+)$ & $(0,0,0,-)$ & $(0,0,1,+)$ & $(0,0,1,-)$ & $(0,1,0,+)$ & $(0,1,0,-)$ & $(0,1,1,+)$ & $(0,1,1,-)$\\
\hline $(0,0,+)$ & $I$ & $\sigma_z\otimes I$ & $I\otimes\sigma_x$
&
$\sigma_z\otimes\sigma_x$ & $I\otimes\sigma_z$ & $\sigma_z\otimes\sigma_z$ & $I\otimes\sigma_z\sigma_x$ & $\sigma_z\otimes\sigma_z\sigma_x$\\
\hline $(0,0,-)$ & $I\otimes\sigma_z$ & $\sigma_z\otimes\sigma_z$
&
$I\otimes\sigma_z\sigma_x$ & $\sigma_z\otimes\sigma_z\sigma_x$ & $I$ & $\sigma_z\otimes I$ & $I\otimes\sigma_x$ & $\sigma_z\otimes\sigma_x$\\
\hline $(0,1,+)$ & $\sigma_z\otimes I$ & $I$
& $\sigma_z\otimes\sigma_x$ & $I\otimes\sigma_x$ & $\sigma_z\otimes\sigma_z$ & $I\otimes\sigma_z$ & $\sigma_z\otimes\sigma_z\sigma_x$ & $I\otimes\sigma_z\sigma_x$\\
\hline $(0,1,-)$ & $\sigma_z\otimes\sigma_z$ & $I\otimes\sigma_z$
& $\sigma_z\otimes\sigma_z\sigma_x$ &
$I\otimes\sigma_z\sigma_x$ & $\sigma_z\otimes I$ & $I$ & $\sigma_z\otimes\sigma_x$ & $I\otimes\sigma_x$\\
\hline $(1,0,+)$ & $\sigma_x\otimes I$ & $\sigma_z\sigma_x\otimes
I$ & $\sigma_x\otimes\sigma_x$ &
$\sigma_z\sigma_x\otimes\sigma_x$ & $\sigma_x\otimes\sigma_z$ & $\sigma_z\sigma_x\otimes\sigma_z$ & $\sigma_x\otimes\sigma_z\sigma_x$ & $\sigma_z\sigma_x\otimes\sigma_z\sigma_x$\\
\hline $(1,0,-)$ & $\sigma_x\otimes\sigma_z$ &
$\sigma_z\sigma_x\otimes\sigma_z$ &
$\sigma_x\otimes\sigma_z\sigma_x$ & $\sigma_z\sigma_x\otimes\sigma_z\sigma_x$ & $\sigma_x\otimes I$ & $\sigma_z\sigma_x\otimes I$ & $\sigma_x\otimes\sigma_x$ & $\sigma_z\sigma_x\otimes\sigma_x$\\
\hline $(1,1,+)$ & $\sigma_z\sigma_x\otimes I$ & $\sigma_x\otimes
I$ &
$\sigma_z\sigma_x\otimes\sigma_x$ & $\sigma_x\otimes\sigma_x$ & $\sigma_z\sigma_x\otimes\sigma_z$ & $\sigma_x\otimes\sigma_z$ & $\sigma_z\sigma_x\otimes\sigma_z\sigma_x$ & $\sigma_x\otimes\sigma_z\sigma_x$\\
\hline $(1,1,-)$ & $\sigma_z\sigma_x\otimes\sigma_z$ &
$\sigma_x\otimes\sigma_z$ &
$\sigma_z\sigma_x\otimes\sigma_z\sigma_x$ &
$\sigma_x\otimes\sigma_z\sigma_x$ & $\sigma_z\sigma_x\otimes I$ & $\sigma_x\otimes I$ & $\sigma_z\sigma_x\otimes\sigma_x$ & $\sigma_x\otimes\sigma_x$\\
\hline\hline
$ $ & $(1,0,0,+)$ & $(1,0,0,-)$ & $(1,0,1,+)$ & $(1,0,1,-)$ & $(1,1,0,+)$ & $(1,1,0,-)$ & $(1,1,1,+)$ & $(1,1,1,-)$\\
\hline $(0,0,+)$ & $I\otimes\sigma_x$ & $\sigma_z\otimes\sigma_x$
& $I$ &
$\sigma_z\otimes I$ & $I\otimes\sigma_z\sigma_x$ & $\sigma_z\otimes\sigma_z\sigma_x$ & $I\otimes\sigma_z$ & $\sigma_z\otimes\sigma_z$\\
\hline $(0,0,-)$ & $I\otimes\sigma_z\sigma_x$ &
$\sigma_z\otimes\sigma_z\sigma_x$ &
$I\otimes\sigma_z$ & $\sigma_z\otimes\sigma_z$ & $I\otimes\sigma_x$ & $\sigma_z\otimes\sigma_x$ & $I$ & $\sigma_z\otimes I$\\
\hline $(0,1,+)$ & $\sigma_z\otimes\sigma_x$ & $I\otimes\sigma_x$
& $\sigma_z\otimes I$ & $I$ & $\sigma_z\otimes\sigma_z\sigma_x$ & $I\otimes\sigma_z\sigma_x$ & $\sigma_z\otimes\sigma_z$ & $I\otimes\sigma_z$\\
\hline $(0,1,-)$ & $\sigma_z\otimes\sigma_z\sigma_x$ &
$I\otimes\sigma_z\sigma_x$ & $\sigma_z\otimes\sigma_z$ &
$I\otimes\sigma_z$ & $\sigma_z\otimes\sigma_x$ & $I\otimes\sigma_x$ & $\sigma_z\otimes I$ & $I$\\
\hline $(1,0,+)$ & $\sigma_x\otimes\sigma_x$ &
$\sigma_z\sigma_x\otimes\sigma_x$ & $\sigma_x\otimes I$ &
$\sigma_z\sigma_x\otimes I$ & $\sigma_x\otimes\sigma_z\sigma_x$ & $\sigma_z\sigma_x\otimes\sigma_z\sigma_x$ & $\sigma_x\otimes\sigma_z$ & $\sigma_z\sigma_x\otimes\sigma_z$\\
\hline $(1,0,-)$ & $\sigma_x\otimes\sigma_z\sigma_x$ &
$\sigma_z\sigma_x\otimes\sigma_z\sigma_x$ &
$\sigma_x\otimes\sigma_z$ & $\sigma_z\sigma_x\otimes\sigma_z$ & $\sigma_x\otimes\sigma_x$ & $\sigma_z\sigma_x\otimes\sigma_x$ & $\sigma_x\otimes I$ & $\sigma_z\sigma_x\otimes I$\\
\hline $(1,1,+)$ & $\sigma_z\sigma_x\otimes\sigma_x$ &
$\sigma_x\otimes\sigma_x$ &
$\sigma_z\sigma_x\otimes I$ & $\sigma_x\otimes I$ & $\sigma_z\sigma_x\otimes\sigma_z\sigma_x$ & $\sigma_x\otimes\sigma_z\sigma_x$ & $\sigma_z\sigma_x\otimes\sigma_z$ & $\sigma_x\otimes\sigma_z$\\
\hline $(1,1,-)$ & $\sigma_z\sigma_x\otimes\sigma_z\sigma_x$ &
$\sigma_x\otimes\sigma_z\sigma_x$ &
$\sigma_z\sigma_x\otimes\sigma_z$ &
$\sigma_x\otimes\sigma_z$ & $\sigma_z\sigma_x\otimes\sigma_x$ & $\sigma_x\otimes\sigma_x$ & $\sigma_z\sigma_x\otimes I$ & $\sigma_x\otimes I$\\
\hline

\end{tabular}
\end{center}
\end{table}
\end{center}
\end{widetext}

\subsection{Toffoli Gate $U_T$}
 \label{ss5}

\begin{figure}
\begin{center}
\includegraphics[width=7cm,angle=0]{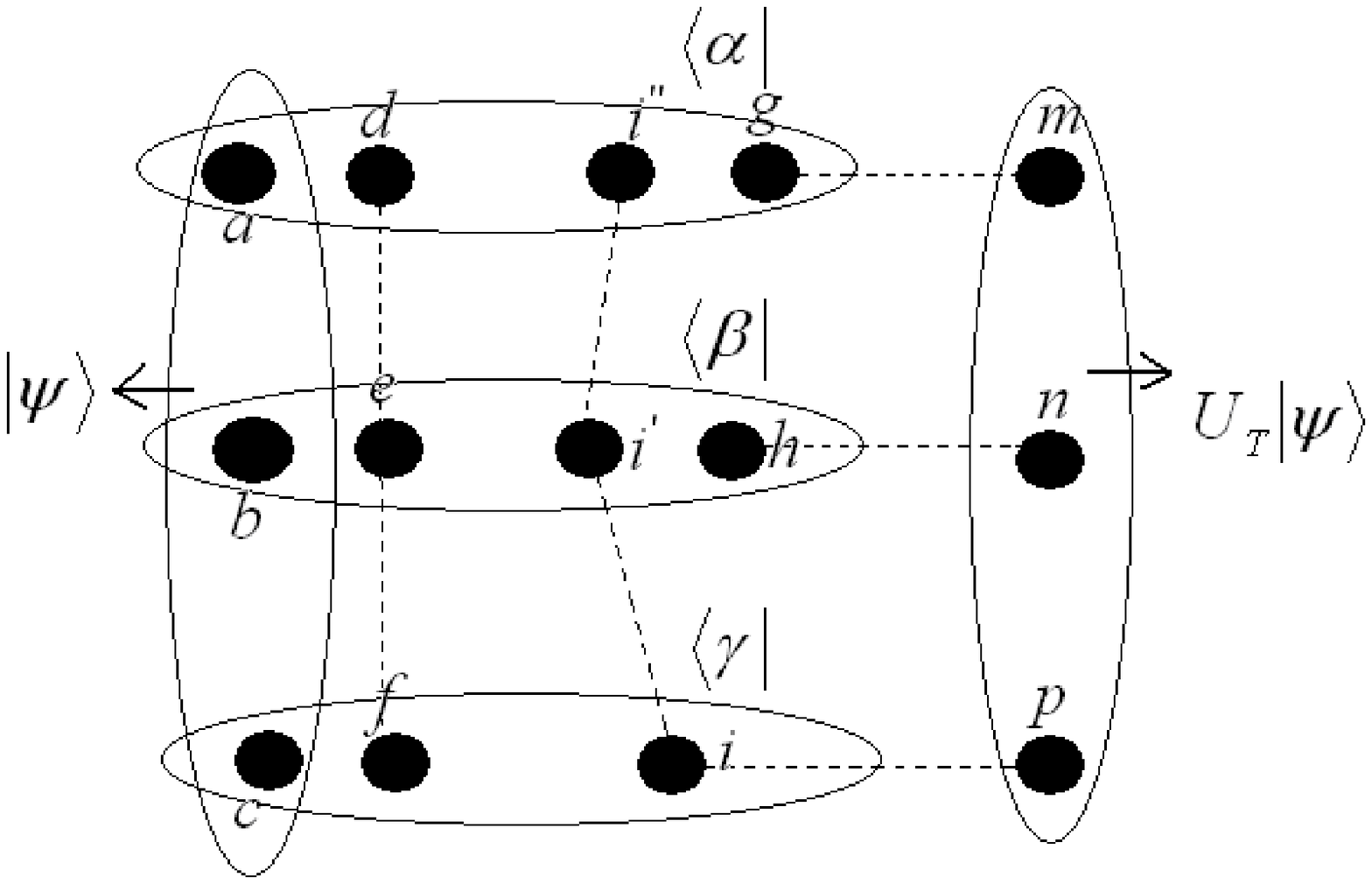}
\end{center}
\caption{ Implementation of Toffoli gate by joint measurement.}
\label{f8}
\end{figure}

 Toffoli gate is the further
development of C-NOT gate. We present the details  for
implementing the Toffoli gate by doing joint measurement as shown
in Fig.\ref{f8}. Explicitly, the entangled state of $def$ is
\begin{eqnarray}
\frac{1}{\sqrt{2}}(|000\rangle+|111\rangle). \label{e38}
\end{eqnarray}
and the entangled state of $gm$, $hn$ are all
\begin{eqnarray}
\frac{1}{\sqrt{2}}(|00\rangle+|11\rangle). \label{e39}
\end{eqnarray}
The entangled state of $i,i^{\prime},i^{\prime\prime},p$ is
\begin{eqnarray}
&&{\sqrt{2}\over
4}(|0_i0_p0_{i^{\prime}}0_{i^{\prime\prime}}\rangle
+|1_i1_p0_{i^{\prime}}0_{i^{\prime\prime}}\rangle
+|0_i0_p1_{i^{\prime}}0_{i^{\prime\prime}}\rangle\nonumber\\
&&-|1_i1_p1_{i^{\prime}}0_{i^{\prime\prime}}\rangle
+|0_i0_p0_{i^{\prime}}1_{i^{\prime\prime}}\rangle+|1_i1_p0_{i^{\prime}}1_{i^{\prime\prime}}\rangle\nonumber\\
&&-|0_i1_p1_{i^{\prime}}1_{i^{\prime\prime}}\rangle
+|1_i0_p1_{i^{\prime}}1_{i^{\prime\prime}}\rangle). \label{e40}
\end{eqnarray}
The joint measurement basis sets are
\begin{eqnarray}
\{|\alpha\rangle\}&=&\{(\sigma_x)^i\otimes(\sigma_z)^j\otimes I\otimes(\sigma_x)^k\nonumber\\
&&(|0_a+_d0_{i^{\prime\prime}}0_g\rangle\pm|1_a-_d0_{i^{\prime\prime}}1_g\rangle)\},
\label{e41}\\
\{|\beta\rangle\}&=&\{(\sigma_x)^i\otimes(\sigma_x)^j\otimes I\otimes(\sigma_x)^k\nonumber\\
&&(|0_b0_e0_{i^{\prime}}0_h\rangle\pm|1_b1_e1_{i^{\prime}}1_h\rangle)\},
\label{e42}\\
\{|\gamma\rangle\}&=&\{(\sigma_x)^i\otimes(\sigma_z)^j\otimes I\nonumber\\
&&(|0_c+_f0_i\rangle\pm|1_c-_f1_i\rangle)\}, \label{e43}
\end{eqnarray}
where
$(i,j,k\in\{0,1\};|\pm\rangle=\frac{1}{\sqrt{2}}(|0\rangle\pm|1\rangle))$.
We note that the extra particles $i$ and $i^{\prime\prime}$ are
necessary in this process.  Some extra operations should be
applied to the wave function of particles $m,n,p$ after joint
measurement too. For instance, as shown in Fig.\ref{f8}, if the
results of measuring-basis sets are
$|\alpha\rangle=|0_a-_d0_{i^{\prime\prime}}1_g\rangle-|1_a+_d1_{i^{\prime\prime}}0_g\rangle$,
$|\beta\rangle=|0_b1_e0_{i^{\prime}}0_h\rangle-|1_b0_e1_{i^{\prime}}1_h\rangle$
and $|\gamma\rangle=|1_c-_f0_i\rangle-|0_c+_f1_i\rangle$, then an
additional operation $\sigma_x\otimes\sigma_z\otimes\sigma_x$ on
particles $m,n,p$ is necessary.

\subsection{ Fredkin (controlled-swap) Gate}
\label{ss6}

\begin{figure}
\begin{center}
\includegraphics[width=7cm,angle=0]{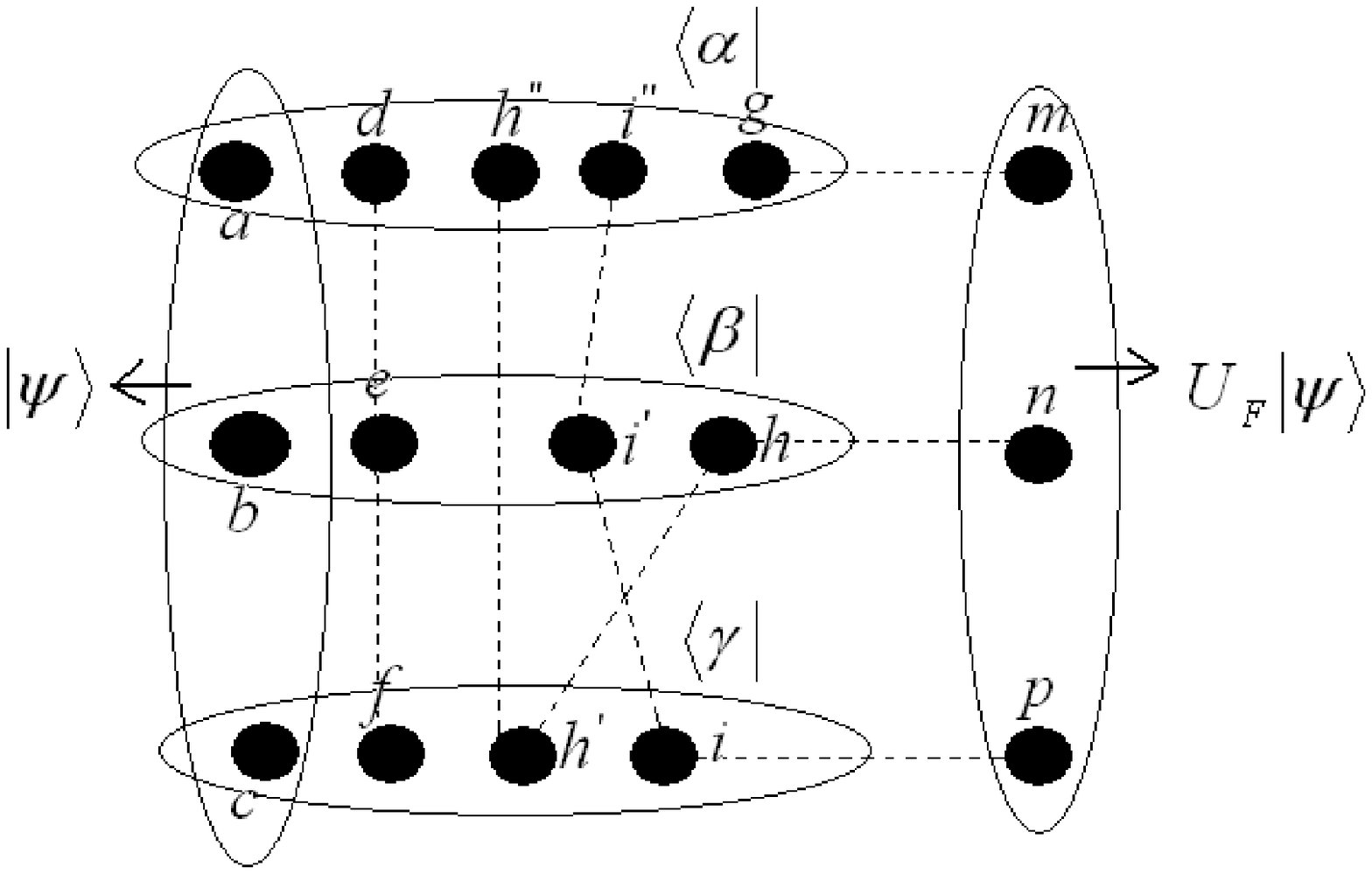}
\end{center}
\caption{ Implementation of Fredkin gate by joint measurement.}
\label{f9}
\end{figure}

Fredkin gate is the development of swap gate in a three-qubit
system. The details of the implementation of this quantum gate is
given in Fig.\ref{f9}. Here the entangled state of $d,e,f$ is
\begin{eqnarray}
\frac{1}{\sqrt{2}}(|000\rangle+|111\rangle). \label{e44}
\end{eqnarray}

The entangled state of $g,m$ is
\begin{eqnarray}
\frac{1}{\sqrt{2}}(|00\rangle+|11\rangle). \label{e45}
\end{eqnarray}

The entangled state of $i,i^{\prime},i^{\prime\prime},p$ is
\begin{eqnarray}
&&\frac{\sqrt{2}}{4}(|0_i0_p0_{i^{\prime}}0_{i^{\prime\prime}}\rangle+|1_i1_p0_{i^{\prime}}0_{i^{\prime\prime}}\rangle
+|0_i0_p1_{i^{\prime}}0_{i^{\prime\prime}}\rangle\nonumber\\
&&-|1_i1_p1_{i^{\prime}}0_{i^{\prime\prime}}\rangle+
|0_i0_p0_{i^{\prime}}1_{i^{\prime\prime}}\rangle-|1_i0_p0_{i^{\prime}}1_{i^{\prime\prime}}\rangle\nonumber\\
&&-|0_i1_p1_{i^{\prime}}1_{i^{\prime\prime}}\rangle+|1_i1_p1_{i^{\prime}}1_{i^{\prime\prime}}\rangle).
\label{e46}
\end{eqnarray}

The entangled state of $h,h^{\prime},h^{\prime\prime},n$:
\begin{eqnarray}
&&\frac{\sqrt{2}}{4}(|0_n0_{h^{\prime}}0_h0_{h^{\prime\prime}}\rangle+|0_n1_{h^{\prime}}0_h0_{h^{\prime\prime}}\rangle
+|0_n0_{h^{\prime}}0_h1_{h^{\prime\prime}}\rangle\nonumber\\
&&-|1_n1_{h^{\prime}}0_h1_{h^{\prime\prime}}\rangle+
|1_n0_{h^{\prime}}1_h0_{h^{\prime\prime}}\rangle+|1_n1_{h^{\prime}}1_h0_{h^{\prime\prime}}\rangle\nonumber\\
&&+|0_n0_{h^{\prime}}1_h1_{h^{\prime\prime}}\rangle+|1_n1_{h^{\prime}}1_h1_{h^{\prime\prime}}\rangle).
\label{e47}
\end{eqnarray}

The measuring-basis sets are
\begin{eqnarray}
\{|\alpha\rangle\}&=&\{(\sigma_x)^i\otimes(\sigma_z)^j\otimes(\sigma_x)^k\otimes I\otimes(\sigma_x)^w\nonumber\\
&&(|0_a+_d0_{h^{\prime\prime}}0_{i^{\prime\prime}}0_g\rangle\pm|1_a-_d1_{h^{\prime\prime}}1_{i^{\prime\prime}}1_g\rangle)\},
\label{e48}\\
\{|\beta\rangle\}&=&\{(\sigma_x)^i\otimes(\sigma_x)^j\otimes I\otimes(\sigma_x)^k\nonumber\\
&&(|0_b0_e0_{i^{\prime}}0_h\rangle\pm|1_b1_e1_{i^{\prime}}1_h\rangle)\},
\label{e49}
\end{eqnarray}
\begin{eqnarray}
\{|\gamma\rangle\}=\{(\sigma_x)^i\otimes(\sigma_z)^j\otimes(\sigma_x)^k\otimes I\nonumber\\
(|0_c+_f0_{h^{\prime}}0_i\rangle\pm|1_c-_f1_{h^{\prime}}1_i\rangle)\},
\label{e50}
\end{eqnarray}
where $(i,j,k,w\in\{0,1\})$. It is worth poiting the extra
particles $h^{\prime}$, $h^{\prime\prime}$, $i^{\prime}$,
$i^{\prime\prime}$ are crucial to implement the Fredkin gate.

\section{Summary}
\label{s4}

 In this paper, we have constructed explicitly the elementary gates
for the measurement-based quantum computation, including the
generalized controlled-Z gate, the phase gate and $\frac{\pi}{8}$
gate, the swap gate, the C-NOT gate, the Fredkin gate and Toffoli
gate. We have studied the relation between the form of the
measuring-basis and the entangled states of the pairs. It is found
that the they have an exquisite relation among them. In some
cases, some parts in the joint measurement basis are necessarily
distinct for different qubits in a group so that they can
implement the desired quantum gate. It is interesting to mention
that as shown in Ref.\cite{vbs}, the teleportation-based quantum
computation and the VBS quantum computation scheme are equivalent.
It is also equivalent to the cluster-state quantum computation. In
addition, we note that the patterns of entangled states in the
process of constructing C-NOT gate and swap gate are the same and
they only differ at the  ways of the joint measurement. For these
matters, the topological construction of quantum
computation\cite{rasetti1,rasetti2} is worthwhile for further
study at this point. In all these schemes, entanglement is the
core of quantum computation. In the cluster-state quantum
computation scheme, the entanglement is condensed into the initial
state, while in the teleportation-based quantum computation as
studied in this paper, the entanglement in injected into the
quantum computing system by using entangled pairs of particles.

This work is supported by the National Fundamental Research
Program, Grant No. 001CB309308, China National Natural Science
Foundation, Grant No. 60073009, 10325521, the SRFDP program of
Education Ministry of China.

\end{document}